\documentclass[prb,preprint,showpacs,showkeys]{revtex4}

\usepackage{hyperref}
\usepackage[english]{babel}
\usepackage[T1]{fontenc}
\usepackage[latin1]{inputenc}
\usepackage[dvips]{graphicx}
\usepackage{amsmath}
\usepackage{array}
\usepackage{bm}
\usepackage{fancyhdr}
\usepackage{makeidx}
\usepackage{subfigure}
\usepackage[v2]{xy}
\usepackage{epsfig}
\usepackage{amsfonts}
\usepackage{amssymb}
\usepackage{color}
\hypersetup{colorlinks=true,breaklinks=true,urlcolor=blue,linkcolor=blue,bookmarksopen=true}
\makeindex
\begin{document}

\title{Spin transfer torques in magnetic tunnel junctions}
\date{\today}
\author{A. Manchon$^1$,
N. Ryzhanova$^{1,2}$, M. Chshiev$^3$, A. Vedyayev$^{1,2}$, K.-J.
Lee$^4$, B. Dieny$^{1}$} \affiliation{$^1$SPINTEC, URA 2512
CEA/CNRS, CEA/Grenoble, 38054 Grenoble Cedex 9, France\\
$^2$Department of Physics, M. V. Lomonosov Moscow State
University, 119899 Moscow, Russia\\ $^3$MINT Center, University of
Alabama, P.O. Box 870209, Tuscaloosa, Alabama, USA\\
$^4$Department of Materials Science and Engineering, Korea
University, Seoul 136-713, Korea} \email{aurelien.manchon@m4x.org}
\begin{abstract}
This chapter presents a review on spin transfer torque in magnetic
tunnel junctions. In the first part, we propose an overview of
experimental and theoretical studies addressing current-induced
magnetization excitations in magnetic tunnel junctions. The most
significant results are presented and the main observable
characteristics are discussed. A description of the mechanism of
spin transfer in ferromagnets is finally proposed. In the second
part, a quantum description of spin transport in magnetic tunnel
junctions with amorphous barrier is developed. The role of
spin-dependent reflections as well as electron incidence and
spin-filtering by the barrier are described. We show that these
mechanisms give rise to specific properties of spin transfer in
tunnel junctions, very different from the case of metallic
spin-valves. In the third part, the theoretical observable
features of spin transfer in magnetic tunnel junctions are derived
and the validity of these results is discussed and compared to
recent experiments. To conclude this chapter, we study the
mechanism of spin transfer in half-metallic tunnel junctions,
expected to mimic MgO-based magnetic tunnel junctions.
\end{abstract}
\keywords{Spin Transfer Torque, Magnetic tunnel junctions,
Tunnelling Magnetoresistance, Current-induced Magnetization
Switching}\maketitle
\begin{small}
 \tableofcontents
\end{small}
\clearpage
\section{Introduction}
The study of the coupling between an electrical current and
localized spins in transition metals, leading to giant
magnetoresistance effects \cite{baibich,binasch}, has renewed our
knowledge of fundamental electronics and opened wide fields of
research in this domain. The idea that a spin-polarized current
may in turn act on the local magnetization of such a ferromagnet
have been proposed in the late 1970's by Berger \cite{berger78},
when investigating the interaction between a domain wall and an
electrical current.\par

However, this torque - usually called spin transfer torque (STT) -
exerted by the spin-polarized current on the local magnetization
requires high current densities which can only be reached in
sub-micronic devices (nano-pillars, point contacts or nano-wires).
The development of thin film deposition techniques, as well as
electronic lithography in the early 1990's led to the fabrication
of spin-valve pillars with dimensions as small as 100$\times$100
nm$^2$. Spin-valves, first studied by Dieny et al. \cite{dieny} in
1991, consist of two ferromagnetic thin layers (less than 10
nm-thick), separated by a metallic (Cu, Al) or tunnelling
(Al$_2$O$_3$, MgO, TaOx) spacer. One of the ferromagnet is pinned
by an antiferromagnetic system so that its magnetization direction
is only weakly affected by an external magnetic field.\par

The theoretical demonstration of spin transfer torque in metallic
spin valves (SVs) ten years ago \cite{slonc96,berger96} gave a new
breath to giant magnetoresistance related studies \cite{GMRreview},
promising exciting new applications in non-volatile memories
technology \cite{applSTT} and radio-frequency oscillators
\cite{rfSTT}. A number of fundamental studies in metallic spin
valves revealed the different properties of spin torque and led to a
deep understanding of current-induced magnetization dynamics
\cite{katine,kiselev,rippard,urazhdin,alhajdarwish}. Particularly,
several theoretical studies described the structure of the torque in
metallic magnetic multilayers and showed the important role of
averaging due to quantum interferences, spin diffusion and spin
accumulation \cite{stiles02,autres,jpcm}.\par

Since the first experimental evidence of spin-dependent tunnelling
\cite{jullieremeservey}, magnetic tunnel junctions (MTJs) have
attracted much attention because of the possibility to obtain large
tunnelling magnetoresistance (TMR) at room temperature
\cite{moodera}. The possibility to use MTJs as sensing elements in
magnetoresistive heads, as non-volatile memory elements or in
reprogrammable logic gates has also stimulated a lot of
technological developments aiming at the optimization of MTJs'
transport properties and their implementation in silicon-based
circuitry \cite{applSTT,ieee}. Because of these applications, MTJs
have been intensively studied and the role of interfaces
\cite{leclair}, barrier \cite{sharma}, disorder \cite{tsymbal98} and
impurities \cite{imp} have been addressed in many publications
\cite{reviewTMR}. The recent achievement of current-induced magnetic
excitations and reversal in MTJs \cite{huai,fuchs} has renewed the
already very important interest of the scientific community in
MTJs.\par

The recent observation of spin transfer torque in low RA (resistance
area product) MTJs using amorphous \cite{huai,fuchs} or crystalline
barriers \cite{ieee,ideka} opened new questions about the transport
mechanism in MTJs with non collinear magnetization orientations. As
a matter of fact, whereas the current-perpendicular-to-plane (CPP)
transport in SVs is mostly diffusive and governed by spin
accumulation and relaxation phenomena \cite{autres,jpcm}, spin
transport in magnetic tunnel junctions is mainly ballistic and
governed by the coupling between spin-dependent interfacial
densities of states: all the potential drop occurs within the tunnel
barrier. The characteristics of spin transfer torque are thus
expected to be strongly different in MTJs compared to SVs.\par

In this chapter, we propose a description of spin transfer torque in
magnetic tunnel junctions, highlighting the differences with
metallic spin valves. In section \ref{s:torques}, an overview of the
experiments on spin transfer torque is given as well as a
description of the origin of STT in arbitrary ferromagnetic
systems.\par In section \ref{s:micro}, the quantum origin of spin
transfer torque in MTJs is described using a simple free-electron
approach. The selection of the incident electrons due to the tunnel
barrier is depicted and the relaxation of the transverse and
longitudinal components of the spin density (spin accumulation) is
discussed. It is shown that these two effects may contribute to a
non negligible field-like term (also called out-of-plane component),
contrary to SVs where this term is negligible.\par

In section \ref{s:macro}, we present the angular and bias
dependencies of the in-plane and out-of-plane components of spin
transfer torque. The important angular asymmetry usually observed in
metallic systems disappears in magnetic tunnel junctions due to the
reduced influence of the longitudinal spin accumulation on the
transverse spin current. Then, in agreement with different theories
and very recent experiments, we show that the bias dependencies of
the two components of STT exhibit non linear variations due to the
specific non linear transport through the tunnel barrier. We also
discuss the existence of other sources which can strongly affect
this bias dependence, such as the existence of interfacial
asymmetry, incomplete absorption of the transverse component of spin
current or, most important, emission of spin waves due to hot
electrons.\par

Finally in section \ref{s:wfhm}, we present the influence of
increasing s-d exchange coupling on spin torque and especially
discuss the case of half metallic tunnel junctions, which might
mimic MgO-based MTJs. In half metallic electrodes, the spin transfer
exponentially decays near the interface still giving rise to a non
zero torque on the local magnetization.

\section{Overview of experiments and models\label{s:torques}}

The observation of spin transfer torque in magnetic tunnel
junctions is only very recent (2004) due to the difficulty to
obtain high-quality low RA MTJs. As a matter of fact, as we
stressed out in the introduction, observing the magnetic influence
of spin transfer torque requires the injection of high current
densities in the MTJs, of the order of 10$^7$A/cm$^2$ while
conserving a high current polarization. Reducing the thickness of
the tunnel barrier generally leads to both the reduction of TMR,
as well as the appearance of pinholes \cite{pinholes} (metallic
conduction channel within the tunnelling barrier). The discovery
of spin-filtering effect through MgO crystalline barrier
\cite{butler,butler2} allowed to obtain low resistance magnetic
tunnel junctions together with high current polarization, thus
fulfilling the requirements for the observation of STT in MTJs.
Diao et al. \cite{diao} and Huai et al. \cite{huai2} have compared
the current-induced magnetization reversal in MgO-based and
AlOx-based MTJ and showed that the effective polarization $p$ of
the interfacial densities of states is significantly higher in
MgO-MTJ ($p\approx$46\%) than in AlOx-MTJ ($p\approx$22\%), due to
spin-filtering effects in crystalline MgO barrier. Even if the
existence of such interfacial polarization is questionable
\cite{discussp,slonc05}, this estimation illustrates the
significant improvement achieved with MgO-based MTJs.\par

\subsection{Current-induced magnetization switching}
\subsubsection{General properties}
\begin{figure}[ht]
    \centering
        \includegraphics[width=10cm]{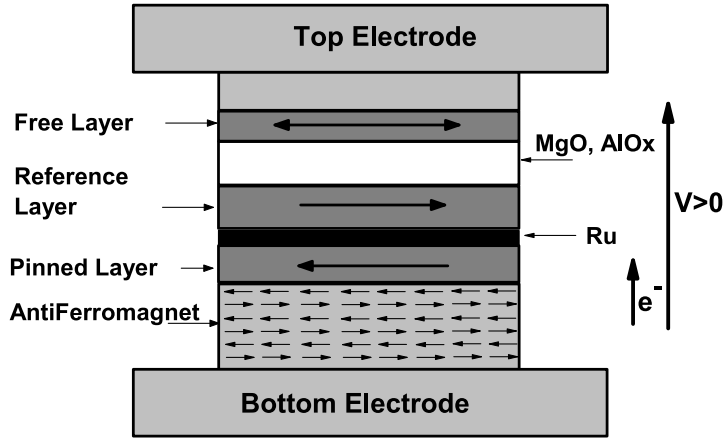}
\caption{Schematics of a magnetic tunnel junction. The bias
voltage is defined positively when the electrons flow from the
reference layer toward the free layer.}
    \label{fig:mtjmgo}
\end{figure}

As we stated in the introduction, a magnetic tunnel junction is a
tunnelling spin valve, as displayed in Fig. \ref{fig:mtjmgo},
composed of two ferromagnetic electrodes (CoFe, CoFeB) separated
by a tunnelling barrier. One ferromagnetic layer (reference layer)
is antiferromagnetically coupled (usually through a thin Ru layer)
to a so-called "pinned layer". This pinned layer is magnetically
coupled to an antiferromagnet (IrMn, FeMn). This technique, known
as synthetic antiferromagnet \cite{saf}, strongly stabilizes the
reference layer while reducing the dipolar field emitted on the
free layer. The free layer magnetization may then be oriented by
an external field, while keeping the magnetization of the
reference layer in a fixed direction.\par
The first observation of current-induced magnetization switching
in magnetic tunnel junctions has been performed by Huai et al.
\cite{huai} and Fuchs et al. \cite{fuchs} in AlOx-based low RA MTJ
(RA<10$\Omega.\mu$m$^2$), in nano-pillar with elliptic shape
(120$\times$230 nm$^2$ in Ref. \cite{huai}).\par

\begin{figure}[ht]
    \centering
        \includegraphics[width=14cm]{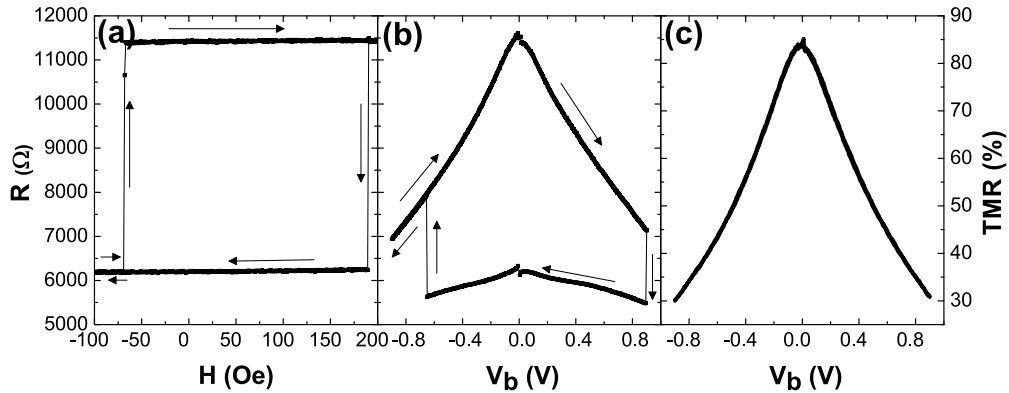}
\caption{Resistance of a CoFeB/MgO/CoFeB MTJs versus (a) the
external field ($V$=10 mV) and (b) the applied bias voltage ($H$=45
Oe). (c) Tunnelling magnetoresistance as a function of the bias
voltage ($H$=45 Oe). TMR$=83.7\%$ and $A=50\times100 nm^2$.}
        \label{fig:RURH}
\end{figure}

The influence of spin transfer torque in magnetic tunnel junctions
is observed by measuring resistance loops as a function of the
external applied field $H$ and the applied bias voltage $V$, as
displayed in Fig. \ref{fig:RURH}. In this figure, we measured the
resistance of a MgO-based MTJ, composed of CoFeB ferromagnetic
electrodes. The resistance loop as a function of the external
field $H$ for a fixed applied bias voltage is given in Fig.
\ref{fig:RURH}(a), while the resistance loop as a function of the
bias voltage $V$ for a fixed external field is given in Fig.
\ref{fig:RURH}(b).\par
One observes sharp resistance jumps in Fig. \ref{fig:RURH}(b) for
positive and negative bias which correspond to the switching of
the free layer magnetization from antiparallel to parallel and
vice-versa, respectively. In this junction, the critical current
needed to switch the free layer magnetization is
5$\times$10$^6$A/cm$^2$. The drop of resistance as a function of
the bias voltage is associated with a drop of TMR (see Fig.
\ref{fig:RURH}(c)). This drop has been attributed to spin-waves
emissions by hot electrons \cite{zhang97} as well as to the
energy-dependence of the density of states at the junction
interfaces. Note that this drop does not exist in metallic spin
valves since only Fermi electrons significantly contribute to the
electrical current in metals.\par

Since these first observations, many efforts have been carried out
in order to obtain low critical current magnetization switching in
MTJs. Dieny et al. \cite{brevD}, Fuchs et al. \cite{fuchs0} and
Huai et al. \cite{huai3} proposed dual type MTJs, in order to
reduce the critical switching current. These structures are of the
type \cite{fuchs0} CoFe$_1$/AlOx/CoFe$_{Free}$/Cu/CoFe$_2$, where
CoFe$_1$ and CoFe$_2$ are antiparallel and the Cu/CoFe$_2$
interface is used to reflect the minority electrons towards
CoFe$_{Free}$ in order to enhance the spin transfer torque in this
layer. With this scheme, critical current were divided by a factor
3.\par

Another method has been proposed by Inokuchi et al. \cite{patent}.
By inserting a non magnetic layer made of Zr, Hf, Rh, Ag, Au or V on
the top of the free layer, it is possible to reduce the critical
current by one order of magnitude and to reach critical current
densities of 5$\times 10^5$ A.cm$^{-2}$.\par

\subsubsection{STT versus TMR\label{s:sttmr}}

\begin{figure}[ht]
    \centering
        \includegraphics[width=10cm]{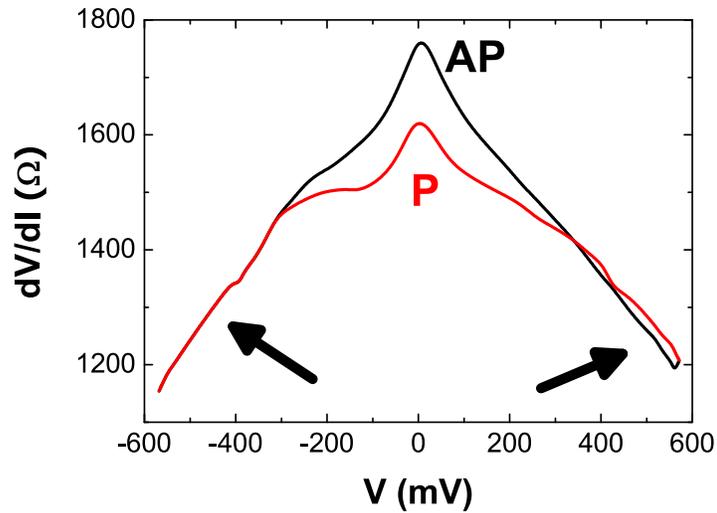}
\caption{Current-induced magnetization switching in AlOx-based
MTJ, measured at 77 K. This switching is associated with a
complete quenching of the TMR. From Ref. \cite{fuchs}.}
    \label{fig:fuchs0}
\end{figure}

An interesting point has been underlined by Fuchs et al.
\cite{fuchs} in their pioneering experiment, when observing
current-induced magnetization switching at 77 K. As displayed on
Fig. \ref{fig:fuchs0}, the magnetization of the free layer could
be switched from antiparallel (black line) to parallel (red line)
by applying an external current. The most interesting is that the
magnetization switching occurred at a bias voltage at which the
TMR was roughly zero, as shown by the arrows on Fig.
\ref{fig:fuchs0}.\par

This experiment demonstrates that the TMR decrease does not prevent
the spin transfer. As a matter of fact, whereas the polarization of
the collecting electrode decreases when increasing the bias voltage
(due to energy-dependence of the interfacial density of states as
well as magnon emission), the polarization of the incident electrons
is only weakly affected. Consequently, a current-induced
magnetization switching may occur although the overall TMR is zero.
In fact, Levy and Fert \cite{swmtj} have shown that the contribution
of hot electrons-induced spin-wave emission may play an important
role in such systems.\par

\subsection{Current-induced magnetization excitations}

Current-induced magnetization excitations are of great interest
for applications, in particular controlling the noise spectrum of
read-head devices or generating hyper-frequencies. However, the
generation of magnetic excitations by a polarized current in MTJs
is rather difficult because of the voltage limitation of the
tunnel barrier which undergo electrical breakdown when submitted
to bias voltage of typically 1 V.\par

A first study of the "spin-diode effect" was published by
Tulapurkar et al. \cite{tula}, in 2005. The authors showed that
the injection of a small radio-frequency ac-current into a
MgO-based MTJ can generate a dc-voltage across the device. This
dc-voltage appears when the frequency of the ac-current is close
to the natural frequency of FMR excitations. This resonance can be
tuned by an external magnetic field. By this way, Tulapurkar et
al. were the first to observe a non negligible "effective field"
term, $b_j$, which was found to be linear as a function of the
bias voltage. Recent developments of this technique were achieved
by Kubota et al. \cite{kubota}. They will be described in section
\ref{s:macro}.\par

Another technique was proposed by Sankey et al.
\cite{sankeyfmr,sankeynat}. By studying the influence of spin
transfer torque on the ferromagnetic resonance of the free layer,
the authors were able to determine the bias dependence of the spin
transfer torque. These results will be described in section
\ref{s:macro}.\par
\begin{figure}[ht]
    \centering
        \includegraphics[width=12cm]{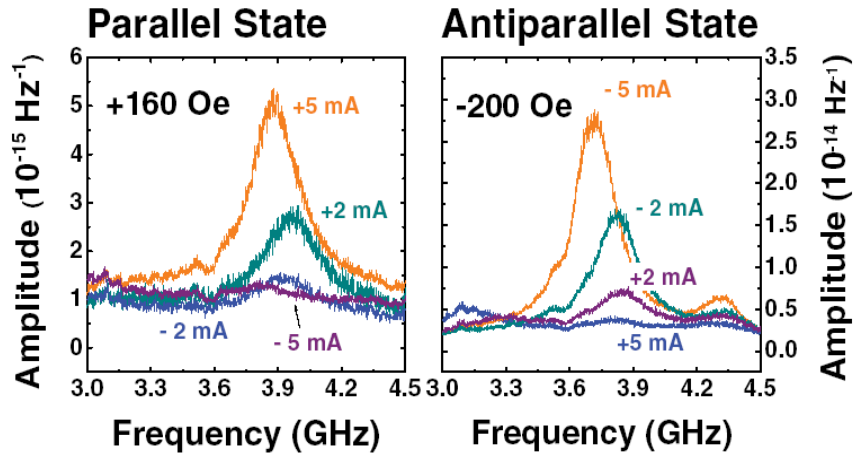}
\caption{Thermally activated FMR spectra of AlOx-based MTJ, as a
function of the injected current in parallel and antiparallel
state. From Ref. \cite{seb}.}
    \label{fig:seb}
\end{figure}
The influence of spin torque on thermally activated ferromagnetic
resonance was also studied \cite{nazarov,zhu1}. Petit et al.
\cite{seb} have demonstrated the influence of spin transfer torque
on thermal noise in MTJs. Fig. \ref{fig:seb} displays the
thermally activated FMR spectra of a AlOx-based MTJ as a function
of the injected current. In parallel configuration, the amplitude
of the FMR peak increases as a function of positive current and
decreases when the injected current is negative (and inversely in
antiparallel configuration). Once again, the authors demonstrated
the strong influence of the $b_j$ term on the magnetization
dynamics.\par

\subsection{Origin of spin transfer torque}
After this short overview on previous relevant experiments, let us
describe the physical origin of spin transfer torque. To do so, we
will proceed in two steps: firstly, a phenomenological description
of spin transfer will be presented, using a simple conceptual
scheme; secondly, the expression of spin transfer torque in an
arbitrary ferromagnet will derived from quantum mechanical
consideration, justifying the phenomenological approach.

\subsubsection{Phenomenological description}
\begin{figure}[ht]
    \centering
        \includegraphics[width=10cm]{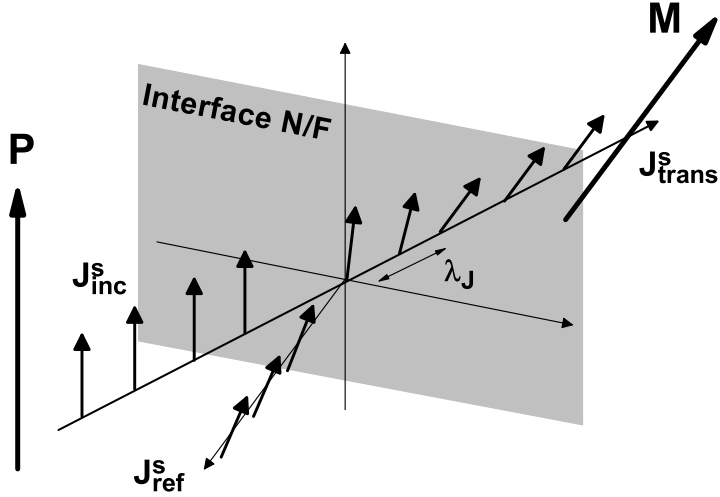}
\caption{Schematics of spin transfer between two magnetic layers.
The polarized electrons flowing from left to right are quickly
reoriented (on a length $\lambda_J$) when arriving in the right
layer. The balance between inward and outward currents is transfer
to the local magnetization.}
    \label{fig:stt_princ}
\end{figure}

The principle of spin transfer between two ferromagnetic layers is
sketched on Fig. \ref{fig:stt_princ}. Let us consider an
electrical current, spin-polarized along the ${\bf P}$ direction
(the electrical current may be polarized by a previous
ferromagnetic layer for example). This spin-polarized current
impinges on a N/F interface, where N  is a normal metal (or a
tunnel barrier) and F is a ferromagnetic metal whose magnetization
${\bf M}$ forms an angle $\theta$ with ${\bf P}$, so that ${\bf
P}.{\bf M}=\cos\theta$ ($\theta\neq0$). Johnson et al. \cite{JS}
and Van Son et al. \cite{vanson} showed that an out-of-equilibrium
magnetization (also called spin accumulation in diffusive systems,
or spin density in ballistic systems) appears at this interface,
due to the different spin-scattering rates in the N and F layers.
In our system, since the impinging current is not polarized
following ${\bf M}$, the rising out-of-equilibrium magnetization
${\bf m}$ possesses three components. It can then exert a torque
on the local magnetization ${\bf M}$ of the form ${\bf
T}=-J_{sd}/\mu_B{\bf M}\times{\bf m}$. Because of the fast angular
precession of the electrons spin around ${\bf M}$ and due to the
relaxation of the spin accumulation ${\bf m}$ in the ferromagnet
F, the transverse component of the spin accumulation is quickly
absorbed close to the N/F interface, on a length scale
$\lambda_J$, usually smaller than 1 nm in metallic spin-valves
\cite{stiles02,urzh}.\par Another way to understand spin transfer
torque is to consider that the electrical current possesses an
initial polarization, described by the spin current ${\bf
J}^s_{inc}$. One part of this impinging current is reflected by
the N/F interface, giving rise to a reflected (backward) spin
current ${\bf J}^s_{ref}$. In the adiabatic regime (the electron
spin precession is fast compared to the local magnetization
dynamics), after a length $\lambda_J$, itinerant electrons are
aligned along the local magnetization ${\bf M}$ and the
transmitted spin current is then ${\bf J}^s_{trans}\neq{\bf
J}^s_{inc}$. The reflected spin current ${\bf J}^s_{ref}$ being
generally small, the net balance of angular moment yields the
transverse component of the incident spin current: ${\bf
J}^s_{inc}-{\bf J}^s_{trans}-{\bf J}^s_{ref}={\bf J}^s_{inc\bot}$
(note that {\it transverse} means {\it transverse to} ${\bf M}$).
Thus, the impinging electrons lose the transverse component of
their magnetic moment which is transmitted to the localized
electrons, responsible for the local magnetization ${\bf M}$. This
spin transfer is translated in a torque of the form:${\bf
T}=-\nabla {\bf J}^s$.\par

Stiles et al. \cite{stiles02} have described the origin of spin
transfer torque at a N/F interface, where N is a metal. The authors
proposed three mechanisms giving rise to spin transfer in ballistic
systems. First, the spin dependence of the interfacial reflection
and transmission coefficients induces a {\it discontinuity} of the
spin current so that one part of the transverse component of spin
current is absorbed at the interface. This discontinuity gives rise
to a torque in the plane (${\bf P}$, ${\bf M}$) which tends to align
${\bf P}$ and ${\bf M}$. Secondly, the {\it spin precession} around
the local magnetization ${\bf M}$, after averaging over the whole
Fermi surface, gives rise to the complete absorption of the
transverse spin current on a length scale of the order of
$\lambda_J=1$ nm. Finally, after reflection by the interface, the
electron spin forms an angle with {\it both} ${\bf P}$ and ${\bf
M}$. This {\it spin rotation} yields the appearance of another
component of the spin torque, perpendicular to the plane (${\bf P}$,
${\bf M}$) and called out-of-plane torque.\par

Thus, these three contributions give rise to a torque exerted by the
spin accumulation on the local magnetization, written as:
\begin{equation}\label{eq:troq}
{\bf T}=a_j {\bf M}\times\left({\bf M}\times {\bf P}\right)+b_j
{\bf M}\times {\bf P}
\end{equation}
where $a_j$ and $b_j$ are the in-plane and out-of-plane torque
amplitudes. Note that in the first theories of spin transfer
torque by Slonczewski \cite{slonc89,slonc96,slonc02} and Berger
\cite{berger96,berger01}, the authors only derived $a_j$ because
they considered that the electron spin remains in the (${\bf P}$,
${\bf M}$) plane, as corroborated by $ab-initio$ calculations
\cite{stiles02}. These theories apply to metallic spin valves
where, due to the small length $\lambda_J$, spin transfer is
assumed to take place very close to the interface \cite{fertstt}.
However, Edwards et al. \cite{edwards} have derived a sizable
out-of-plane torque in metallic spin-valves using non equilibrium
Green's functions and interestingly, Zhang et al. \cite{zlf} have
demonstrated that taking into account the spin precession in the
transport model significantly enhances the $b_j$ term. In magnetic
tunnel junctions, both $a_j$ and $b_j$ term arise from different
mechanisms that will be described in section \ref{s:micro}.\par

Injecting the spin transfer torque ${\bf T}$ in the
Landau-Lifshitz-Gilbert (LLG) equation, one obtains the modified LLG
equation, describing the magnetization dynamics of the free layer,
submitted to both an external field and a spin-polarized electrical
current:

\begin{equation}\label{e:llg}
\frac{\partial{\bf M}}{\partial t}=-\gamma{\bf M}\times\left({\bf
H_{eff}}+b_j{\bf P}\right)+\alpha{\bf M}\times\frac{\partial{\bf
M}}{\partial t}-\gamma a_j{\bf M}\times\left({\bf M}\times{\bf
P}\right)
\end{equation}
where $\gamma$ is the gyromagnetic ratio, $\alpha$ is the Gilbert
damping and ${\bf H_{eff}}$ is the effective field, including the
anisotropy field, the demagnetizing field and the external applied
field. From Eq. \ref{e:llg}, the out-of-plane torque acts as an
effective field while the in-plane torque acts as an effective
(anti-)damping. As a function of its sign, $a_j$ may excite or
damp magnetic excitations in the magnetization ${\bf M}$, whereas
$b_j$ only affects the energy surface of the ferromagnetic layer.
Different magnetic behavior may be observed: magnetization
switching from a stable state to another, stabilization of
magnetic states at low energy minima, or magnetic excitations
(coherent and incoherent precessions).\par

\subsubsection{Spin transfer in an arbitrary ferromagnet\label{s:sttdescr}}

All along this section, we consider the {\it s-d} model in which two
populations of electrons coexist: itinerant electrons ({\it sp}-type
or itinerant {\it d}-type electrons) and localized electrons ({\it
d}-type mainly). The localized electrons give rise to the local
magnetization of the ferromagnet. We also assume that the {\it d}
local moments remain stationary. This model applies to the
electronic structure of ferromagnetic electrodes whose compositions
lie on the negative slope side of the Slater-Néel-Pauling curve
\cite{stearns73} (Ni, Co, NiFe, CoFe).

\paragraph{Itinerant electrons dynamics}\par\medskip

The motion of itinerant electrons in the ferromagnetic materials
are represented by the non-relativistic single electron
Hamiltonian including $s-d$ coupling:
\begin{equation}\label{e:sdh}
H=\frac{p^2}{2m}+U(\textbf{r})-J_{sd}(\bm{\sigma}.\bm{S}_d)
\end{equation}
where the first and second terms are the kinetic and potential
energies, while the third term is the $s-d$ exchange energy,
$\bm{S}_d$ being the unit vector of the local magnetization due to
the localized electrons and $J_{sd}$ the {\it s-d} exchange
constant. Let us define the local spin density
$\textbf{m}(\textbf{r},t)$ and the local spin current density of
itinerant electrons $\textbf{J}_s$ as
\begin{equation}
\textbf{m}(\textbf{r},t)=\Psi^*(\textbf{r},t)\frac{\hbar}{2}\bm{\sigma}\Psi(\textbf{r},t)\\
\end{equation}
\begin{equation}
\textbf{J}_s=-\frac{\hbar^2}{2m}Im\{\Psi^*(\textbf{r},t)\bm{\sigma}\nabla_\textbf{r}\Psi(\textbf{r},t)\}
\end{equation}
and the temporal derivative of the spin density is:
\begin{equation}\label{e:spincont}
\frac{d}{dt}\textbf{m}(\textbf{r},t)=\frac{\hbar}{2}\{\frac{d}{dt}\Psi^*\bm{\sigma}\Psi+\Psi^*\bm{\sigma}\frac{d}{dt}\Psi\}
\end{equation}
where $\Psi=\left(\Psi^\uparrow,\Psi^\downarrow\right)$ is an
arbitrary 2-dimension Hartree-Fock wave function. The two
dimensions refer to up ($\uparrow$) and down ($\downarrow$) spin
projection of the Hartree-Fock wave function.\par From the
time-dependent Schrodinger equation $i\hbar d\Psi/dt=H\Psi$, we
obtain the spin density continuity equation:

\begin{equation}\label{e:spincont4} \frac{d\textbf{m}}{dt}=-\nabla \textbf{J}_s+
\frac{2J_{sd}}{\hbar}\bm{S_{d}}\times\textbf{m}
\end{equation}
To correctly describe the ferromagnetic system under
consideration, one should add the interactions between electrons
and lattice, for example. In diffusive regime, one can introduce a
spin relaxation term which depends on the spin density
\cite{zhangli} $\Gamma(\textbf{m})=\frac{\textbf{m}}{\tau_{sf}}$:

\begin{equation}\label{e:spincont5}
\frac{d\textbf{m}}{dt}=-\nabla \textbf{J}_s+
\frac{2J_{sd}}{\hbar}\bm{S_d}\times\textbf{m}-\frac{\textbf{m}}{\tau_{sf}}
\end{equation}

Eqs. \ref{e:spincont4} and \ref{e:spincont5} are of great
importance to understand the role of spin transport in STT. One
can see that the temporal variation of the spin density (or spin
accumulation) arises from the contribution of three sources: the
spatial variation of spin current density, the torque exerted by
the background magnetization and a scattering source which acts as
a spin sink.\par

\paragraph{Localized electrons dynamics}\par\medskip

The Hamiltonian of a single localized spin submitted to a time
dependent external field and to an external current flow is:
\begin{equation}
H=-\frac{g\mu_B}{\hbar}\bm{S_d}.\bm{B}-\frac{2J_{sd}}{\hbar}\bm{S_d}.\textbf{m}=-\frac{g\mu_B}{\hbar}\bm{S_d}.\bm{B^{eff}}
\end{equation}
where $g$ is the Lande factor, $\mu_B$ is the Bohr magnetron,
$\bm{S_d}$ is the localized spin, $\bm{B}$ is the external
magnetic field, $\textbf{m}$ is the out-of-equilibrium spin
density of the itinerant electrons and $\bm{B^{eff}}$ is the
effective field due to the combination between the external field
and the itinerant electron spin density. Applying Ehrenfest
theorem \cite{miltat} leads to
\begin{equation}
\frac{d<\bm{S}>}{dt}=-\frac{g\mu_B}{\hbar}<\bm{S}>\times\bm{B}^{eff}
\end{equation}
where <> denotes averaging over all the localized states,
$<\bm{S}>=\bm{S_d}$. We can rewrite this equation as:
\begin{equation}
\frac{d\bm{S_d}}{dt}=-\frac{g\mu_B}{\hbar}\bm{S_d}\times\bm{B}-\frac{2J_{sd}}{\hbar}\bm{S_d}\times\textbf{m}
\end{equation}
The first term includes all the interactions with magnetic fields,
like external field, magnetocrystalline anisotropy. The second
term arises from the presence of itinerant electrons. In order to
take into account the damping of the localized spin, one has to
consider a more complete Hamiltonian that includes many body
interactions which leads to the usual Landau-Lifshitz-Gilbert
equation:
\begin{equation}\label{e:sdh1}
\frac{d\bm{S_d}}{dt}=-\frac{g\mu_B}{\hbar}\bm{S_d}\times\bm{B}-\frac{2J_{sd}}{\hbar}\bm{S_d}\times\textbf{m}+\alpha\bm{S_d}\times\frac{d\bm{S_d}}{dt}
\end{equation}
where $\alpha$ is the phenomenological Gilbert damping
coefficient.

\paragraph{Modified LLG dynamic equation}\par\medskip

Averaging Eq. \ref{e:sdh1} over all the electrons of the structure
and setting $g=2$, and $\gamma=2\mu_B/\hbar$, we obtain the
modified LLG equation:

\begin{equation}\label{e:sdh2}
\frac{d\textbf{M}}{dt}=-\gamma\textbf{M}\times{\bf
H}_{eff}-\gamma\frac{J_{sd}}{\mu_B}\textbf{M}\times\textbf{m}+\alpha\textbf{M}\times\frac{d\textbf{M}}{dt}
\end{equation}

Here \textbf{M} is the local magnetization, \textbf{m} is the
out-of-equilibrium spin accumulation or spin density of itinerant
electrons, and \begin{equation}
\bm{H}_{eff}=\frac{H_KM_x}{M_s}\bm{e}_x+\frac{2A_{ex}}{M_s^2}\nabla^2\textbf{m}-4\pi
M_z\bm{e}_z+H_{ext}\bm{e}_x
\end{equation}
where $H_K$ is the anisotropy field, $A_{ex}$ is the exchange
constant, and $4\pi M_z$ is the demagnetization field. The term
proportional to $J_{sd}$ is a torque exerted by the spin
accumulation $\textbf{m}$ on the local magnetization $\textbf{M}$,
similar to the one given in Eq. \ref{e:spincont5}. It is
interesting to note that only the transverse spin accumulation
$\textbf{m}$ has an influence on the background magnetization
state in the form of a torque $\bm{T}$ along two axes:
\begin{equation}\label{eq:LLG1}
\bm{T}=-\frac{J_{sd}}{\mu_B}\textbf{M}\times\textbf{m}=-\frac{J_{sd}}{\mu_B}\left[m_x\textbf{M}\times\bm{P}-m_y\textbf{M}\times\left(\textbf{M}\times\bm{P}\right)\right]
\end{equation}
where $\textbf{P}$ is the unit vector parallel to the
magnetization of the pinned layer and $\textbf{M}$ is the unit
vector parallel to the magnetization of the free layer. The first
term in the right-hand-side of Eq. \ref{eq:LLG1} is called the
field-like term (or out-of-plane torque, or current-induced
interlayer exchange coupling) and the second term is the usual
Slonczewski term (or in-plane torque).\par The time scale of
itinerant spins dynamics is two orders of magnitude shorter than
the time scale of the background magnetization dynamics. So one
can consider, in a first approximation, that the itinerant spins
can be described by the steady state equation (see Eqs.
\ref{e:spincont4} and \ref{e:spincont5}):
\begin{eqnarray}
\label{e:spincont7}
&&-\nabla\textbf{J}_s(\textbf{r},t)=\frac{2J_{sd}}{\hbar}\textbf{m}\times\textbf{M}\;\;\text{(ballistic
system)}\\\label{e:spincont8}
&&-\nabla\textbf{J}_s(\textbf{r},t)=\frac{2J_{sd}}{\hbar}\textbf{m}\times\textbf{M}+\frac{\textbf{m}}{l_{sf}}\;\;\text{(diffusive
system)}
\end{eqnarray}
Eqs. \ref{e:spincont7}-\ref{e:spincont8} imply that the spatial
transfer of spin density per unit of time from the itinerant
$s$-electrons to the localized $d$-electrons (left-hand side
terms) is equivalent to a torque exerted by the transverse spin
accumulation on the local magnetization (right-hand side terms),
modulated by the relaxation of the spin accumulation in diffusive
regime.

\subsection{Theories of spin transfer in magnetic tunnel junctions}

Slonczewski first proposed a free electron model of spin transport
in a MTJ with an amorphous barrier \cite{slonc89}, deriving TMR,
in-plane spin transfer torque and zero bias interlayer exchange
coupling (IEC). This first model only considered electrons at Fermi
energy, neglecting all non-linear tunnel behaviors (consequently,
the out-of-plane torque was found to be zero). In a two band model,
the torque was written as:
\begin{equation}\label{e:slonc}
\bm{T}=\frac{e\kappa^3(\kappa^4-k_\uparrow^2k_\downarrow^2)(k_\uparrow^2-k_\downarrow^2)}{2\pi^2d(\kappa^2+k_\uparrow^2)^2(\kappa^2+k_\downarrow^2)^2}e^{-2\kappa
d}V\textbf{M}\times(\textbf{M}\times\bm{P})
\end{equation}
where $\kappa$ is the barrier wave vector,
$k_{\uparrow,\downarrow}$ are the Fermi wave vectors for majority
and minority spins, $d$ is the barrier thickness and $V$, the bias
voltage across the junction. Note that this model is restricted to
rectangular barrier, so very low bias voltage. More recently,
combining Bardeen Transfer Matrix formalism (BTM) and his previous
results on the relation between torques and spin currents
\cite{slonc02}, the author proposed a more general formula for
in-plane torque in magnetic tunnel junctions
\cite{slonc05,slonc07}:
\begin{eqnarray}
&&\bm{T}=\frac{\hbar}{4}\left[\Gamma_{++}+\Gamma_{+-}-\Gamma_{--}-\Gamma_{-+}\right]\textbf{m}\times(\textbf{m}\times\bm{P})\label{e:torquesl}\\
&&\Gamma_{\sigma\sigma'}=\frac{2\pi eV}{\hbar}\sum_{p,q}\gamma^2_{p,\sigma;q,\sigma'}\\
&&\gamma_{p,\sigma;q,\sigma'}=\frac{-\hbar^2}{2m}\int
dydz(\psi_{p,\sigma}\partial_x
\phi_{q,\sigma'}-\phi_{q,\sigma'}\partial_x \psi_{p,\sigma})
\end{eqnarray}
where $\psi$ and $\phi$ are the orbital wave functions for right
and left interface. This relation stands for electrons whose
energy is close to the Fermi energy. The author underlined
interestingly that Eq. \ref{e:torquesl} may be simplified if the
integrals $\Gamma_{\sigma\sigma'}$ can be separated in the form:
\begin{equation}
\Gamma_{\sigma\sigma'}\propto D_{L,\sigma}D_{R,\sigma'}
\end{equation}
where $D_{L(R),\sigma}$ is the density of states at the left (right)
interface, for spin projection $\sigma$. In this case, it is
straightforward to see that the torque exerted on the right layer is
reduced to:
\begin{equation}
\bm{T}_R=\frac{\hbar}{4}P_L\textbf{M}\times(\textbf{M}\times\bm{P})\label{e:torques2}
\end{equation}
where $P_L$ is the interfacial polarization of the density of
states, as defined by Julliere \cite{jullieremeservey}. This leads
to a bias asymmetry of the spin transfer torque, since the
polarization $P_L$ is bias dependent for only one direction of the
applied voltage. The condition of this separability has been
discussed by Slonczewski \cite{slonc05}, Belaschenko et al.
\cite{belashchenko} and Mathon et al. \cite{umerski1}. These
authors have suggested that the phase decoherences, induced by
disorder in realistic junctions, could reduce the polarization
factors to a product between the interfacial densities of states.
It seems that this assumption is valid in magnetic tunnel
junctions with not so thin barriers, especially in amorphous
AlOx-based MTJs.\par

Theodonis et al. \cite{theo,kalitsov} recently presented a
tight-binding model (TB) of MTJs, taking into account more
realistic band structures than the usual free electron model.
These studies showed that the in-plane torque should present an
important bias asymmetry while the out-of-plane torque should be
of the same order of magnitude with a quadratic dependence on the
bias voltage. This is in agreement with recent studies of
Wilczynski et al. \cite{barnasstt} and Manchon et al.
\cite{manchon}, based on free electron model, as discussed in this
chapter.\par The role of magnons have been addressed by Levy et
al. \cite{swmtj} and by Li et al. \cite{prlli}. It was shown that
magnons emission may strongly influence the bias dependence of
spin transfer torque contributing to modify the absorption length
$\lambda_J$. This mechanism will be discussed in section
\ref{s:macro}.\par

Finally, note that all these theories assume amorphous barriers
and a plane wave description of the transport, although most of
the experiments are carried out on crystalline MgO-based MTJs. A
recent publication from Heiliger et al. \cite{heiliger} addresses
the characteristics of spin transfer torque in Fe/MgO/Fe
crystalline junctions. The dominant contribution of $\Delta_1$
symmetry strongly influences spin torque feature.

\section{Quantum origin of spin torque in magnetic tunnel junctions\label{s:micro}}
We will now describe the spin transport in magnetic tunnel
junctions. Although most of the experiments are nowadays performed
in crystalline MgO-based MTJ, one can get a first insight of TMR and
spin torque by simply considering a free electron model of magnetic
tunnel junctions.\par

We first introduce the free-electron model, and then depict the spin
transport in a MTJ with non-collinear magnetization directions.
Afterward, we will describe the role of the barrier on the spin
transfer torque. Finally, the origin of the torques and coupling
between the two ferromagnetic layers will be explained.
\subsection{Free electron model}

\begin{figure}
    \centering
        \includegraphics{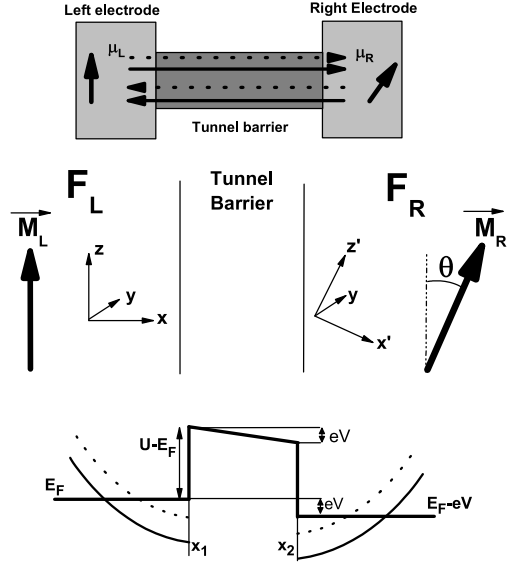}
\caption{Schematics of the magnetic tunnel junction with non
collinear magnetization orientations. Top panel: spin-dependent
out-of-equilibrium transport in a conductor linking two reservoirs
$F_L$ and $F_R$ (whose electrochemical potentials are respectively
$\mu_L$ and $\mu_R$) with non collinear magnetization orientations.
The solid arrows represent the majority spins and the dotted arrows
represent the minority spins. Middle panel: MTJ with non collinear
magnetization orientations. Bottom panel: Corresponding energy
profile of the MTJ. In free-electron approximation, the local
density of states are parabolic for majority (solid line) and
minority (dotted line) electrons with a splitting between the two
spin sub-bands equals to the exchange interaction $J_{sd}$.}
\label{fig:figbase}\end{figure}

The basis of our calculation is depicted in the top panel of Fig.
\ref{fig:figbase}. The out-of-equilibrium magnetic tunnel junction
is modeled by a "conductor" (in the sense that the tunnel barrier
is not infinite) linking two magnetic reservoirs ($F_L$ and $F_R$)
with non collinear magnetizations and with different chemical
potentials $\mu_L$ and $\mu_R$ \cite{datta} ($\mu_L>\mu_R$). A
bias voltage $V=(\mu_L-\mu_R)/e$ is applied across this
"conductor". One has to consider all electrons with majority spins
(solid arrows) and minority spins (dotted arrows), originated from
left (rightward arrows) and right electrodes (leftward arrows). In
low bias limit ($\mu_L\approx\mu_R$), the charge transport can be
approximately determined by the electrons originated only from the
left electrode with an energy between $E_F$ and $E_F-eV$.\par In
our case (middle panel of Fig. \ref{fig:figbase}), the magnetic
tunnel junction is composed of two ferromagnetic layers, $F_L$ and
$F_R$ (made of the same material, for simplicity), respectively
connected to the left and right reservoirs and separated by an
amorphous tunnel barrier. The {\it x}-axis is perpendicular to the
plane of the layers and the magnetization of $F_L$ is oriented
following z: $\textbf{M}_L=M_L\bm{z}$. The magnetization
$\textbf{M}_R$ of $F_R$ is in the ({\it x,z}) plane and tilted
from $\textbf{M}_L$ by an angle $\theta$. In this configuration,
the spin density in the ferromagnetic layer possesses three
components : $\textbf{m}=(m_x,m_y,m_z)$. In $F_L$ (we obtain the
same results considering $F_R$), the transverse components are
$m_x=<\sigma^x>$ and $m_y=<\sigma^y>$, where $\sigma^i$ are the
Pauli spin matrices and <> denotes averaging over orbital states
and spin states, i.e. averaging over electrons energy $E$,
transverse momentum $\bm{\kappa}$ and spin states. The transverse
spin density in the left layer is then given by
$<\sigma^+>=<\sigma^x+i\sigma^y>$ :
\begin{eqnarray}\label{e:1}
&& m_x+im_y=<\sigma^+>=2<\Psi^{*\uparrow}\Psi^{\downarrow}>
\end{eqnarray}
In other words, the in-plane torque is given by the imaginary part
of $<\sigma^+>$, while the out-of-plane torque is given by its real
part. One can understand the product
$<\Psi^{*\uparrow}\Psi^{\downarrow}>$ as a correlation function
between the two projections of the spin of the impinging electrons.
In ballistic regime, the spin of an electron impinging on a
ferromagnet with a spin polarization tilted from the background
magnetization precesses around this magnetization
\cite{stiles02,kalitsov}. Locally, its two projections $\uparrow$
and $\downarrow$ following the quantization axis (defined by the
background magnetization) are then non-zero. As a result, the
electron contributes locally to the transverse spin density $m_x$
and $m_y$. If the electron spin is fully polarized parallel or
antiparallel to this magnetization, no precession occurs and its
contribution to the transverse spin density is zero.\par

We remind that we defined majority (minority) states as the spin
projection parallel (antiparallel) to the magnetization of the left
electrode. Therefore, $<\Psi^{*\uparrow}\Psi^{\downarrow}>$ is the
fraction of electrons whose spin is following $x$ (real part) and
$y$ (imaginary part) in spin space.

In Keldysh out-of-equilibrium formalism \cite{datta,keldysh}, the
conductivity is calculated considering the contribution of the
electrons originating from the left reservoir {\it and} from the
right reservoir (top panel of Fig. \ref{fig:figbase}). The
out-of-equilibrium Green function $G(\bm{r},t,\bm{r}',t')$ (or
Keldysh Green function) is defined as a superposition of these two
contributions:

\begin{equation}\label{e:gmp0}
G\left(\bm{r},t,\bm{r}',t'\right)=f_L\Psi_L\left(\bm{r},t\right)\Psi_L^{*}\left(\bm{r}',t'\right)+f_R\Psi_R\left(\bm{r},t\right)\Psi_R^{*}\left(\bm{r}',t'\right)
\end{equation}
where $\Psi_{L(R)}\left(\bm{r},t\right)$ are the electron wave
functions originating from the left (right) reservoir at the
location $\bm{r}$ and time $t$ and $f_{L(R)}$ are the Fermi
distribution functions in the left and right reservoirs.\par

Thus, the Schrodinger equation of the magnetic tunnel junction is:

\begin{equation}\label{e:2}
    H\Psi=\left(\frac{p^2}{2m}+U-J_{sd}\left(\bm{\sigma}.\bm{S_d}\right)\right)\left(
\begin{tabular}{c}
$\Psi^{\uparrow}$ \\
$\Psi^{\downarrow}$
\end{tabular}\right)=E\left(\begin{tabular}{c}
$\Psi^{\uparrow}$ \\
$\Psi^{\downarrow}$
\end{tabular}\right)
\end{equation}
where $\bm{\sigma}$ the vector in Pauli matrices space :
$\bm{\sigma}=(\sigma^x, \sigma^y, \sigma^z)^T$, $E$ is the
electron energy, $U$ is the spin-independent potential along the
junction:

$$J_{sd}\left(\bm{\sigma}.\bm{S_d}\right)=J_{sd}\sigma^z\ \ \ \mbox{and}\ \ \ U=E_F\ \ \ \mbox{for}\ \ \ x<x_1$$
$$J_{sd}\left(\bm{\sigma}.\bm{S_d}\right)=0\ \ \ \mbox{and}\ \ \ U(x)=U_0-\frac{x-x_1}{x_2-x_1}eV\ \ \ \mbox{for}\ \ \ x_1<x<x_2$$
$$J_{sd}\left(\bm{\sigma}.\bm{S_d}\right)=J_{sd}\left(\sigma^z\cos\theta+\sigma^x\sin\theta\right)\ \ \ \mbox{and}\ \ \ U=E_F-eV\ \ \ \mbox{for}\ \ \ x>x_2$$

We consider that the potential drop occurs essentially within the
barrier and we assume the bias voltage is low compared to the
barrier height ($V<<U/e$). This allows to use WKB approximation to
determine the wave functions inside the barrier. Furthermore, the
free electron approximation implies parabolic dispersion laws which
also restricts our study to low bias voltage.\par

In the 2-dimensional Hartree-Fock representation, spin-dependent
current and spin density are defined using the out-of-equilibrium
lesser Keldysh Green function:

\begin{eqnarray}\label{e:gmp1}
G_{\sigma\sigma'}^{-+}\left(\bm{r},\bm{r}'\right)=&&\int d\epsilon \left(f_L\left[\Psi_L^{\sigma'\left(\uparrow\right)*}\left(\bm{r}'\right)\Psi_L^{\sigma\left(\uparrow\right)}\left(\bm{r}\right)+\Psi_L^{\sigma'\left(\downarrow\right)*}\left(\bm{r}'\right)\Psi_L^{\sigma\left(\downarrow\right)}\left(\bm{r}\right)\right]\right.\nonumber\\
&&\left.+f_R\left[\Psi_R^{\sigma'\left(\uparrow\right)*}\left(\bm{r}'\right)\Psi_R^{\sigma\left(\uparrow\right)}\left(\bm{r}\right)+\Psi_R^{\sigma'\left(\downarrow\right)*}\left(\bm{r}'\right)\Psi_R^{\sigma\left(\downarrow\right)}\left(\bm{r}\right)\right]\right)
\end{eqnarray}
where $f_L=f^0(\epsilon)$, $f_R=f^0(\epsilon+eV)$, and
$f^0(\epsilon)$ is the Fermi distribution at 0 K. In-plane
($a_{j}$$\textbf{M}\times(\textbf{M}\times\bm{P})$) and
out-of-plane torques ($b_j\textbf{M}\times\bm{P}$) can now be
determined from Eq. \ref{e:1}, whereas spin-dependent electrical
current densities are calculated from the usual local definition:
\begin{eqnarray}\label{e:defiecstt}
&&b_j+ia_j=\frac{J_{sd}}{\mu_B}<\sigma^+>=2\frac{J_{sd}}{\mu_B}\frac{a_0^3}{(2\pi)^2}\int\int
G^{-+}_{\uparrow\downarrow}(x,x,\epsilon)\kappa d\kappa
d\epsilon\\\label{eq:mz}
&&m_z=\frac{J_{sd}}{\mu_B}\frac{a_0^3}{(2\pi)^2}\int\int
\left[G^{-+}_{\uparrow\uparrow}(x,x,\epsilon)-G^{-+}_{\downarrow\downarrow}(x,x,\epsilon)\right]\kappa
d\kappa d\epsilon\\\label{e:defcurrent}
&&J_{\uparrow\left(\downarrow\right)}=\frac{\hbar e}{4\pi m_e}\int\int\left[\frac{\partial}{\partial x}-\frac{\partial}{\partial x'}\right]G^{-+}_{\uparrow\uparrow(\downarrow\downarrow)}(x,x',\epsilon)|_{x=x'}\kappa d\kappa d\epsilon\\
&&J=J_\uparrow+J_\downarrow
\end{eqnarray}
$G^{-+}_{\uparrow\uparrow}(x,x,\epsilon)$ and
$G^{-+}_{\downarrow\downarrow}(x,x,\epsilon)$ are the
energy-resolved local density-of-states (LDOS) for up- and
down-spins respectively, whereas $\int
G^{-+}_{\uparrow\uparrow}(x,x,\epsilon)d\epsilon$ and $\int
G^{-+}_{\downarrow\downarrow}(x,x,\epsilon)d\epsilon$ give the
density of up- and down-electrons at location $x$ along the
structure.\par

To illustrate the above calculation, we use material parameters
adapted to the case of Co/Al$_2$O$_3$/Co structure: the Fermi wave
vectors for majority and minority spins are respectively
$k_F^{\uparrow}=1.1$ \AA$^{-1}$, $k_F^{\downarrow}=0.6$
\AA$^{-1}$, the barrier height is $U-E_F=1.6$ eV, the effective
electron mass within the insulator is $m_{eff}$=0.4
\cite{bratkovsky} and the barrier thickness is $d$=0.6 nm. These
parameters have been choosen to fit the experimental I-V
characteristics of the magnetic tunnel junctions studied in Ref.
\cite{seb}. In all this section, the magnetizations form an angle
of $\theta$=90°. We will justify this choice in the following.\par

\subsection{Spin transport in a MTJ}

\begin{figure}
    \centering
        \includegraphics[width=8cm]{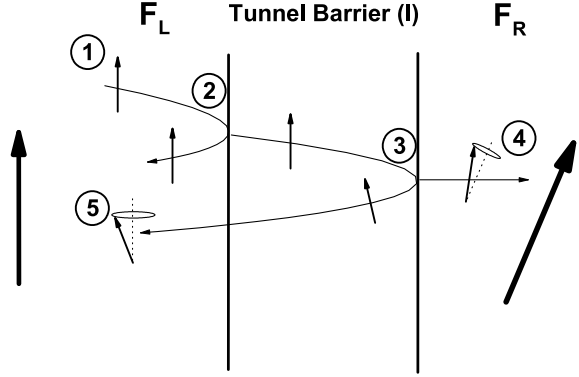}
\caption{Schematics of the principle of spin transport in a
magnetic trilayer with non collinear electrodes magnetizations.
Step 1: the electron spin is polarized along the magnetization of
the left electrode. Step 2: After the first
reflection/transmission by $F_L/I$ interface the reflected and
transmitted parts remain in a pure spin state. Step 3: The
reflection/transmission by the second interface $I/F_R$ reorients
the electron spin. Step 4 and 5: The transmitted and reflected
spins precess around the local magnetization.} \label{fig:fig2}
\end{figure}

Although spin-dependent tunnelling is a well known process, the
description we give here is of great importance to understand the
specific characteristics of spin transfer torques in tunnelling
transport. In this part, we will consider the linear approximation
in which the bias voltage $V_b$ is low enough so that the current
is due to Fermi electrons injected from the left electrode. When
the electrodes magnetizations are non collinear, the electrons are
no more described as pure spin states, but as a mixing between
majority and minority states. For example, let us consider one
electron from the left reservoir, initially in majority spin
state, impinging on the right electrode (see Fig. \ref{fig:fig2} -
step 1). The first reflection (step 2) at the $F_L/I$ interface do
not introduce any mixing since the insulator is non magnetic.
However, when (the transmitted part of) this electron is reflected
or transmitted by the second interface $I/F_R$ (step 3), the
resulting state in the right electrode is a mixing between
majority and minority states since the quantization axis in the
right electrode is different from the quantization axis in the
left electrode. Then, the transmitted spin is reoriented and
precesses (step 4) around the magnetization of the right
electrode. Furthermore, the reflected electron (step 5) is also in
a mixed spin state and precesses around the left electrode
magnetization. In other words, after transport through the
barrier, the electron spin is reflected/transmitted with an angle.
This reorientation gives rise to spin transfer torque.\par

Note that there is no reason why the electron spin should remain in
the plane of the electrodes magnetization. We will see that after
the reorientation, the electron spin possesses three components in
spin space (and so two transverse components).\par

\subsection{Incidence selection in an amorphous barrier}

\subsubsection{$\kappa$-selection due to tunnelling}

It is well know that in non magnetic tunnel junctions, the
transmission of an impinging electrons dependent on its incident
direction. As a matter of fact, the effective barrier thickness
involved in the tunnelling process is larger for grazing incidence
than for normal incidence. The transmission coefficient decreases
exponentially with the in-plane wave vector $\kappa$, so that only
electrons whose wave vector is close to the perpendicular
incidence significantly contribute to the tunnelling
transport.\par
\begin{figure}
\centering
        \includegraphics[width=8cm]{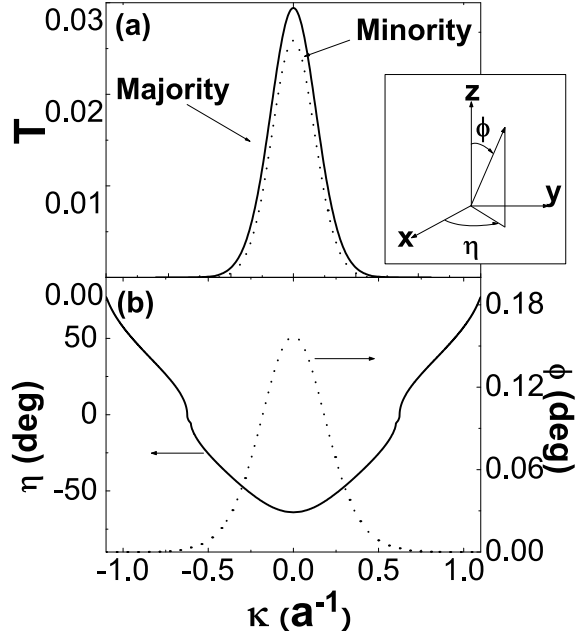}
\caption{(a) Reflectivity of initially majority (solid line) and
minority (dotted line) electrons as a function of the in-plane
wave vector; (b) reflection angles $\eta$ (solid line) and $\phi$
(dotted line) of an initially majority as a function of $\kappa$.
The applied bias voltage is $V_b=0.1$ V and $\theta$=90$^\circ$.
Insert: definition of the reflections angles.} \label{fig:filtre}
\end{figure}
Furthermore, in magnetic tunnel junctions, the transmission
coefficients also depend on the spin projection of the electrons,
as well as on the magnetic configuration of the ferromagnetic
electrodes. This "$\kappa$-selection" is illustrated in Fig.
\ref{fig:filtre}(a). As discussed previously, when the electrodes
magnetization are non-collinear, the spin of an impinging
electron, originally in a pure spin state, is reoriented after
reflection so that the reflected state is in a mixed spin state.
In our case, only the reflection coefficients of the conserved
spin part are reported in Fig. \ref{fig:filtre}(a).\par

Note that only a very small part of the injected polarized wave is
flipped during the tunnelling process. However, this does not mean
that spin transfer torque is small in MTJs, since only {\it
coherent} mixed states contribute to the transverse spin density,
which is responsible of the spin transfer torque.\par

\subsubsection{Spin selection due to ferromagnets}

Following the previous discussion about spin reorientation (see
Fig. \ref{fig:fig2}), it is possible to deduce the angles at which
the electron spin is reflected by the barrier. We define the
azimuthal angle $\eta$ and the polar angle $\phi$ as indicated in
the insert of Fig. \ref{fig:filtre}(a).\par Fig.
\ref{fig:filtre}(b) displays these angles as a function of the
in-plane wave vector $\kappa$. The azimuthal angle $\eta$ varies
between -64$^\circ$ to +77$^\circ$ while the polar angle $\phi$
remains very small (less than 0.2$^\circ$, which means that the
electron spin stays very close to the quantization axis, as
discussed above). At $\kappa=0.6$ \AA$^{-1}$ (corresponding to
$k_F^{\downarrow}$), $\eta=0$ which indicates that the effective
spin density lies in the plane of the magnetizations
$\left(\textbf{M}_L,\textbf{M}_R\right)$. Finally, the polar angle
does not vary with the distance, which means that the reflected
electron spin precesses around $O_z$ with a small angle $\phi$. A
"bulk" spin transfer results from the interferences of all the
reflected electrons.\par

The strong dependence of $\eta$ as a function of the in-plane wave
vector $\kappa$, combined with the $\kappa$-selection close to the
normal incidence (see Fig. \ref{fig:filtre}(a)), implies that the
effective spin of the transmitted electrons possesses an important
out-of-plane component. In other words, the effect of the
spin-dependent tunnelling is to strongly enhance the out-of-plane
component of the spin torque, compared to metallic spin valves. As
a matter of fact, in metallic spin-valves, the whole Fermi surface
contributes to the spin transport so that the effective angle
$\eta$ is very small \cite{stiles02} and correlatively the
out-of-plane torque is negligible.\par

\begin{figure}
    \centering
        \includegraphics[width=10cm]{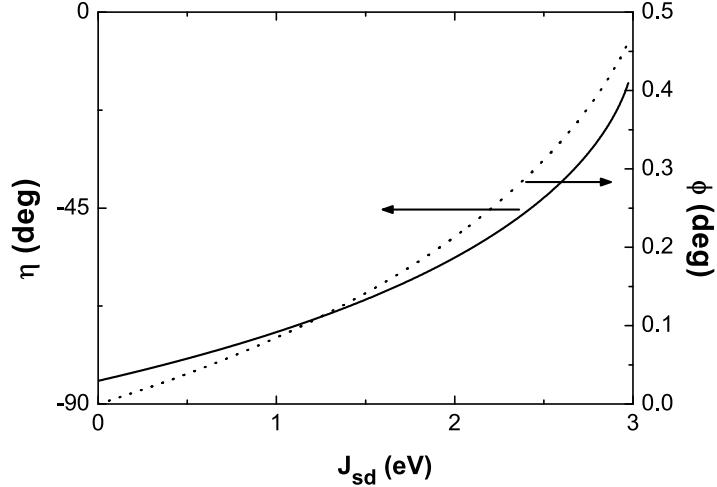}
\caption{Reflection angles as a function of the {\it s-d} exchange
constant, for a Fermi electron initially in majority spin state.
The parameters are the same as in Fig. \ref{fig:filtre}.}
\label{fig:anglespin_Jsd}\end{figure}

Fig. \ref{fig:anglespin_Jsd} shows the dependence of the angles as
a function of the {\it s-d} exchange constant $J_{sd}$ for
perpendicular incidence $\kappa=0$. Quite intuitively, the
precession angle $\phi$ increases with $J_{sd}$ whereas the
initial azimuthal angle $\eta$ decreases in absolute value with
$J_{sd}$. The spin-filtering effect (the selection between
majority and minority spin during the reflection process)
increases with $J_{sd}$ so that the reflected spin direction gets
closer to the plane of the magnetizations.

\subsection{Spin filtering in crystalline structures}

Besides the two fundamental tunnelling selection mechanisms
discussed above, an additional spin filtering mechanism was
proposed by Butler et al.~\cite{butler,butler2} which takes
advantage from the electronic structure of both electrode and
insulator crystalline materials comprising MTJ. It is based on the
fact that only electrons of certain wave function symmetries can
easily propagate through the barrier. For instance, in Fe(001)
only the majority spin channel has electronic states with
$\Delta_1$ symmetry at the Fermi level which in it turn includes
$s$-like character in it. On another hand, the same $\Delta_1$
band in MgO(001) forms an evanescent state in the MgO gap with the
smallest decay rate~\cite{butler,butler2}. As a result,
Fe|MgO|Fe(001) tunnel junction has a very large conductance in
parallel state due to fairly transparent $\Delta_1$ majority
channel at $k_{||}=0$. Antiparallel magnetizations configuration,
on a contrary, is low conductive since the $\Delta_1$ symmetry
states does not exist in the minority band structure around the
Fermi level~\cite{butler,butler2}.\par

Spin transfer torque is nowadays usually observed in MgO-based
crystalline junctions, whereas only few theoretical work has been
done on spin transfer in crystalline structures. The first
theoretical studies of Heiliger et al. \cite{heiliger} on
MgO-based MTJs indicate a dominate contribution of the $\Delta_1$
symmetry on spin transport which may affect the observable
characteristics of STT, as discussed in section \ref{s:macro}.

\subsection{Torques and coupling}
The mechanisms we previously described are at the origin of
spin-dependent plane waves in the MTJ. The interferences between
these waves give rise to an out-of-equilibrium magnetization
$\textbf{m}$ which couples the ferromagnetic electrodes.\par In
the linear regime under consideration, the three components of
spin density in the left electrode can be described as follows:
\begin{eqnarray}\label{e:m}
m_{xL}^{\uparrow}+im_{yL}^{\uparrow}=A(V)\sin\theta\left(e^{i(k_1+k_2)(x-x_1)}-r_1^\uparrow e^{-i(k_1-k_2)(x-x_1)}\right)\\
m_{xL}^{\downarrow}+im_{yL}^{\downarrow}=A^*(V)\sin\theta\left(e^{-i(k_1+k_2)(x-x_1)}-r_1^{\downarrow*}e^{-i(k_1-k_2)(x-x_1)}\right)\\
m_{zL}^{\uparrow}=B^\uparrow(V)-\frac{1}{k_1}\left(r_1^{*\uparrow}e^{2ik_1(x-x_1)}+r_1^{\uparrow}e^{-2ik_1(x-x_1)}\right)\\\label{e:m3}
m_{zL}^{\downarrow}=B^\downarrow(V)+\frac{1}{k_2}\left(r_1^{*\downarrow}e^{2ik_2(x-x_1)}+r_1^{\downarrow}e^{-2ik_2(x-x_1)}\right)
\end{eqnarray}
where $A(V)$, $B^{\uparrow,\downarrow}(V)$ and
$r_1^{\uparrow,\downarrow}$ are coefficients depending on the
junction parameters and on the bias voltage \cite{jpcm} and
$k_{1,2}$ are the wave vectors of majority and minority spin,
respectively.\par

Considering $m_{+L}^{\uparrow\left(\downarrow\right)}$ in Eqs.
\ref{e:m}-\ref{e:m3}, two components can be distinguished : the
first one is proportional to $e^{\pm i(k_1+k_2)(x-x_1)}$, and due
to the interference between the incident wave with majority (resp.
minority) spin and the reflected wave with minority (resp.
majority) spin; the second one is proportional to
$e^{-i(k_1-k_2)(x-x_1)}$ and due to the interference between the
reflected waves with majority and minority spins. We note that the
first components of $m_{+L}^{\uparrow}$ and $m_{+L}^{\downarrow}$
are complex conjugated so that their sum is real. Then, the
interference between the incident wave with majority spin and the
reflected wave with minority spin does not contribute to in-plane
torque but only to out-of-plane torque. In-plane torque is then
generated by the coherent interferences between reflected
electrons with opposite spin projection ($\propto
e^{-i(k_1-k_2)(x-x_1)}$).\par

Concerning $m_{zL}$, it is composed of one component proportional
to $e^{\pm 2ik_1(x-x_1)}$, one component proportional to $e^{\pm
2ik_2(x-x_1)}$ and one constant as a function of $x$. The two
formers are due to the interference between waves having the same
spin projection but with opposite propagation direction while the
latter is due to interference between waves having the same spin
projection and the same propagation direction.\par

\begin{figure}
    \centering
        \includegraphics[width=7cm]{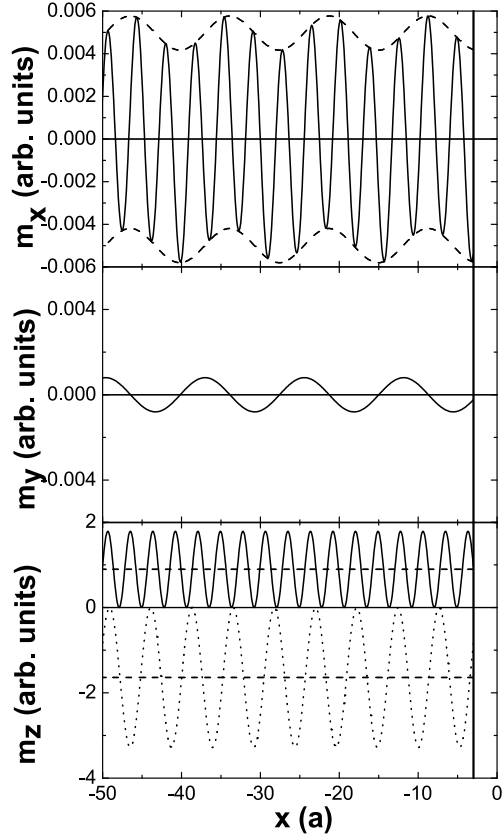}
\caption{Projections of spin density due to Fermi electrons in
perpendicular incidence from the left electrode, as a function of
the distance from the interface. Top panel: $m_x$ component of
spin density (solid line); the dashed lines are the envelopes of
the curve. Middle panel: $m_y$ component of spin density. Bottom
panel: $m_z$ component of spin density due to initially majority
(solid line) and minority (dotted line) spin projection; the
dashed lines are the mean values of the oscillations. The applied
bias voltage is $V_b=0.1$ V. The vertical line on the right is the
interface between the left electrode and the tunnel barrier.}
\label{fig:torqueleft}\end{figure}

Fig. \ref{fig:torqueleft} displays the details of the spin density
components $m_x$, $m_y$ et $m_z$ (described in Eq. \ref{e:m}) in
the left electrode as a function of $x$, when $V_b=0.1$ V. $m_x$
possesses a quite complex behavior with two periods of oscillation
(the dashed lines show the envelope of the curve), whereas $m_y$
is reduced to a single oscillation (The oscillation period
$k_1+k_2$ vanishes when summing the contribution of majority and
minority spins); $m_z$ oscillates around mean values represented
by horizontal dashed lines.\par

Note that the conservative part of the out-of-plane torque
(interlayer exchange coupling at zero bias \cite{jfv,tiusanjpcm})
is only proportional to $e^{-i(k_1+k_2)(x-x_1)}$. But at non zero
bias, the dissipative part of the out-of-plane torque is
proportional to both $e^{-i(k_1+k_2)(x-x_1)}$ and
$e^{-i(k_1-k_2)(x-x_1)}$.\par

\section{Observable properties\label{s:macro}}
Up to now, in order to describe the quantum origin of spin torque in
MTJ, we focused on Fermi electrons and low bias voltage. To depict
the observable properties of spin transfer torque in MTJ, we should
take into account all the electrons from the left and the right
electrodes so as to include non-linear processes.
\subsection{Angular dependence}

\begin{figure}
    \centering
        \includegraphics[width=10cm]{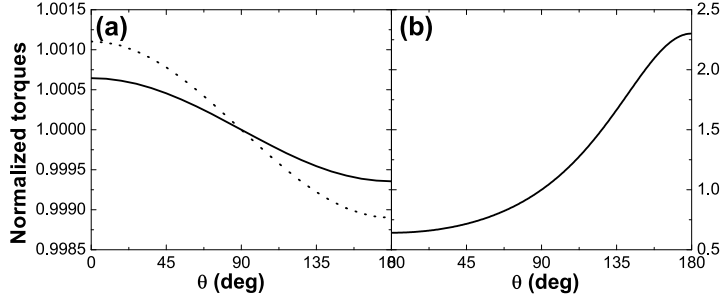}
\caption{(a) Angular dependence of normalized in-plane (solid
line) and out-of-plane torque (dotted line) in a magnetic tunnel
junction; (b) Angular dependence of normalized in-plane torque in
a metallic spin-valve. From Ref. \cite{jpcm}.}
\label{fig:theta}\end{figure}

Fig. \ref{fig:theta}(a) shows the normalized in-plane and
out-of-plane components, $a_j^{norm}$ and $b_j^{norm}$, as a
function of the angle $\theta$ between the electrodes
magnetizations, at $V_b=0$ and $V_b=0.1$ V. The normalized torques
are defined as:
$$\bm{T^{norm}}=\bm{T}/\bm{T(90^\circ)}\sin\theta$$ It clearly
appears that both components are proportional to $\sin\theta$ (the
deviation from $\sin\theta$ is smaller than 10$^{-4}$). This
dependence is strongly different from what was predicted in
metallic spin valves \cite{autres,jpcm,slonc02} (see Fig.
\ref{fig:theta}(b)) and has been attributed \cite{slonc05} to the
{\it single-electron} nature of tunnelling.\par

As a matter of fact, in metallic spin-valves, the spin
accumulation, due to spin-dependent scattering at the interfaces,
modifies the potential profile seen by the electrons. This effect
is due to the multi-electrons nature of diffusive transport, since
the transport of one electron spin is affected by the spin
accumulation rising from the whole spin polarized current. This
spin accumulation strongly influences the angular dependence of
the stack resistance and spin transfer torque \cite{jpcm}.\par

On the contrary, in magnetic tunnel junctions, because of the
important height of the tunnel barrier ($\approx0.8-3.3$ eV), all
the potential drop occurs inside the insulator and the spin
accumulation (i.e. the feedback of the current-induced
longitudinal spin density on the spin current) is negligible. In
this case, the angular dependence of torque is determined by the
angular dependence of the transmission matrix, as discussed in
Ref. \cite{slonc05} and yields a sine shape. In the following, we
will estimate the spin density for $\theta=\pi/2$.\par Note that,
at zero bias, the out-of-plane torque is still non-zero, contrary
to in-plane torque. The conservative part of the out-of-plane
torque (interlayer exchange coupling at zero bias) comes from the
contribution of electrons located under the Fermi level
\cite{jfv,tiusanjpcm}. At zero bias, the currents from left and
right electrodes are equal, but the electron propagation still
corresponds to the scheme shown in Fig. \ref{fig:fig2}: the mixing
between majority and minority states induces a transverse
component in the spin density.\par

\subsection{Decay length of spin density}
As discussed in section \ref{s:sttdescr}, spin transfer torque is
estimated from the transverse component of the spin density. This
spin density (or spin accumulation in diffusive systems) usually
decays due to quantum interferences or spin-dependent scattering, so
that spin torque is generally assumed to be an interfacial
phenomenon.
\subsubsection{Ballistic interferences}
\begin{figure}
    \centering
        \includegraphics[width=6cm]{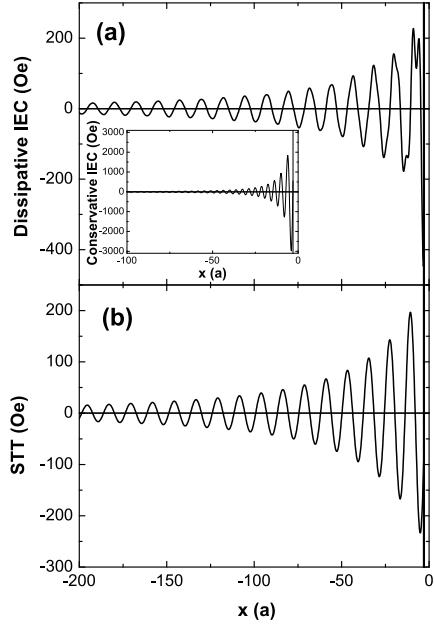}
\caption{Total spin density as a function of the location in the
left electrode: (a) In-plane spin density - inset: In-plane spin
density at zero bias voltage; (b) Out-of-plane spin density. These
quantities are calculated at $V_b=0.1$ V.}
\label{fig:torques_z}\end{figure}

In the present model, no spin-diffusion is taken into account and
the Fermi surface is assumed spherical. Fig. \ref{fig:torques_z}
displays the two components of transverse spin density as a
function of the location in the left electrode. The interference
process between polarized electrons yields a damped oscillation of
the in-plane component $m_x$ (giving rise to the out-of-plane
torque) as presented in Fig. \ref{fig:torques_z}(a). We can
distinguish two periods of oscillation
$T_1=2\pi/\left(k_F^\uparrow-k_F^\downarrow\right)$ and
$T_2=2\pi/\left(k_F^\uparrow+k_F^\downarrow\right)$ whereas at
zero bias, only $T_2$ appears (see inset of Fig.
\ref{fig:torques_z}(a)). This can be easily understood by
considering electrons from left and right electrodes. The
transverse spin density in the {\it left} electrode due to
electrons from the {\it right} electrode is:
\begin{eqnarray}\label{e:m2}
m_{+R}^{\uparrow} =C^\uparrow(V)\sin\theta e^{-i(k_1-k_2)(x-x_1)}\\
m_{+R}^{\downarrow} =C^\downarrow(V)\sin\theta
e^{-i(k_1-k_2)(x-x_1)}
\end{eqnarray}
where $C^{\uparrow,\downarrow}(V)$ are coefficients depending on
the junction parameters and on the bias voltage \cite{jpcm}. It is
now possible to show that in the general expression of transverse
spin density
$$m_+=m_x+im_y=m_{+L}^{\uparrow}+m_{+L}^{\downarrow}+m_{+R}^{\uparrow}+m_{+R}^{\downarrow}$$
the terms proportional to $e^{-i(k_1-k_2)(x-x_1)}$ vanish at zero
bias due to the cancellation of contribution of electrons from the
left and right reservoirs at zero bias voltage
($A(0)+A^*(0)=C^{\uparrow}(0)+C^{\downarrow}(0)$) so that $m_+$
reduces to terms proportional to $e^{\pm i(k_1+k_2)(x-x_1)}$
\cite{manchon}. Furthermore, these last terms only give a real
component since, as discussed above, the majority and minority
components of $m_y$ (giving rise to the in-plane torque)
compensate each other. Consequently, at zero bias, only the
conservative part of the out-of-plane torque (zero bias interlayer
exchange coupling) exists, due to the interference between
incident and reflected electrons with opposite spin projection
\cite{jfv,tiusanjpcm}. But when the bias voltage is non zero, the
transport becomes asymmetric and the terms proportional to
$e^{-i(k_1-k_2)(x-x_1)}$ do not compensate each other anymore
which leads to two periods of oscillations as shown in Fig.
\ref{fig:torques_z}(a).\par

In-plane component of spin transfer torque, proportional to $m_y$,
exits only at non zero bias and possesses only one period of
oscillation $T_1$ (see Fig. \ref{fig:torques_z}(b)). It is worthy
to note that the transverse components of spin density is damped
by 50\% within the first nanometers, and that the amplitude of the
out-of-plane torque is of the same order than the in-plane torque.
This decay length is very large compared to previous theoretical
predictions \cite{stiles02,slonc02} and experimental
investigations on SV \cite{urzh}. As a matter of fact, the
ballistic assumption holds for distance smaller than the mean free
path ($\approx 5$ nm in Co). In realistic devices, spin diffusion
processes should increase the decay of the transverse components
of spin density.\par

\begin{figure}
    \centering
        \includegraphics[width=8cm]{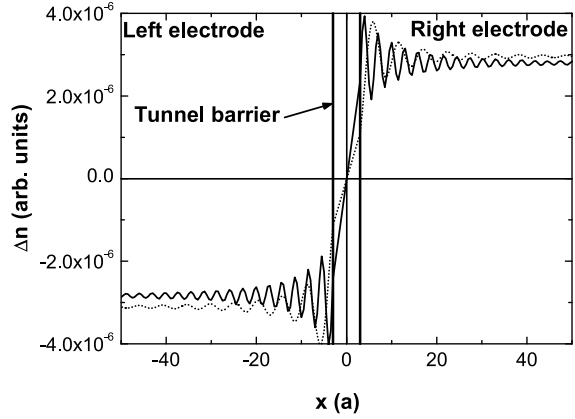}
\caption{Out-of-equilibrium longitudinal spin density throughout the
magnetic tunnel junction for majority (solid line) and minority
(dotted line) electron spin projections. The bias voltage is
$V_b=0.1$ V.}
    \label{fig:mz_z}
\end{figure}
Finally, Fig. \ref{fig:mz_z} shows the out-of-equilibrium
longitudinal spin density $\Delta n$ defined as $\Delta
n^{\uparrow\left(\downarrow\right)}=n^{\uparrow\left(\downarrow\right)}(V_b=0.1)-n^{\uparrow\left(\downarrow\right)}(V_b=0)$.
$\Delta n$ oscillates and asymptotically reaches a non zero value.
This means that when the bias voltage is turned on, a non
equilibrium spin accumulation builds up. However, this effective
spin accumulation is very small ($\Delta n^{\uparrow}-\Delta
n^{\downarrow}\approx 10^{-7}$ electron/atom) and cannot influence
spin current building. Therefore, neglecting the role of
longitudinal spin accumulation (spin density) in MTJ is
justified.\par

\subsubsection{Spin scattering mechanisms}

In real magnetic tunnel junctions, one should take into account
spin-flip processes induced by spin-orbit coupling as well as hot
electrons-induced spin-waves emissions that occur within the
diffusive ferromagnetic electrodes. Spin-orbit induced spin-flip
scattering (Elliott-Yafet mechanism \cite{elliott,yafet}) as well
as spin-wave scattering \cite{fert69} lead to spin-diffusion
length, $l_{sf}$ of 15-30 nm in usual ferromagnetic
electrodes\cite{bass}. This spin-flip should increase the spatial
decay rate of the spin density by a factor of $e^{-l_{sf}x}$.\par

Spin-flip scattering by hot-electrons induced spin wave is a
spin-flip mechanism that specifically occurs in magnetic tunnel
junctions \cite{zhang97}. In tunnel junctions, at non zero bias,
spin-polarized electrons from the left electrode impinge to the
right electrode with an energy higher than the local Fermi energy:
they are called "hot electrons". These hot electrons relax towards
the Fermi level by inelastic scattering involving phonon and
magnon emission. Following the Fermi Golden rule, this spin-waves
emission increases with temperature and energy of the hot
electrons. Li et al. \cite{prlli} have shown that the
spin-diffusion length due to this mechanism is written:
\begin{equation}
l_{sf}\propto J_FE_F/J_{sd}^2V_b
\end{equation}
where $J_F$ is the ferromagnetic exchange constant and $E_F$ the
Fermi energy. The authors find a spin-diffusion length of about
0.5-2 nm for reasonable parameters. This demonstrates the essential
role of magnon emissions in magnetic tunnel junctions.

\subsubsection{Real Fermi surfaces}

In order to more accurately describe spin-dependent transport
throughout crystalline barriers \cite{butler,butler2} (in
particular MgO-based MTJs), the role of defaults in the barrier
\cite{tsymbal}, or interfacial states effects, it is necessary to
go beyond the free electron model and consider the real band
structure of the stack.\par

First principle studies of realistic Co/Cu interfaces \cite{zwier}
(so, metallic spin-valves) showed that the mismatch of the
electronic structure at the interface for spin down electrons
strongly reduces the transverse component of spin density. As a
matter of fact, the spin-dependent transmission at the interface
becomes more complex. In particular, the electron phase
distribution becomes broad and asymmetric \cite{stiles02}. This
leads to a rapid interfacial decay of the transverse spin
accumulation in metallic spin-valves. In MTJ, the non spherical
nature of the spin-dependent Fermi surface
\cite{belashchenko,butler,tsymbal} should also dramatically alter
the transverse spin density. This could explain the fact that the
amplitude of spin torque in the free-electron model we proposed is
two orders of magnitude higher than in experiments.\par

Heiliger et al. \cite{heiliger} recently studied the spin transfer
torque in Fe/MgO/Fe crystalline tunnel junction. The authors
showed that the interfacial spin density decay is even stronger in
this type of MTJ than in metallic spin-valves. This decay is
attributed to the dominant contribution of $\Delta_1$ electrons
for which Fe behaves as a half-metal with respect to this
symmetry. Spin transfer torque arising from the interferences
between majority (propagative states) and minority (evanescent
states) electrons, is localized close to the MgO/Fe interface.
This point will be addressed in more details in section
\ref{s:wfhm}.

\subsection{Bias dependence}
\subsubsection{Free electron model}

The bias dependence of in-plane and out-of-plane torques in MTJ
also presents strong differences with metallic spin-valves. We
first calculate the total spin torque exerted on the left
electrode. Following the definition of Ref. \cite{slonc96} and
Ref. \cite{theo}, the total torque is:
\begin{equation}\label{eq:js}
    \overrightarrow{T}_{total}=\int_{x_1}^{-\infty}-\nabla \bm{J}^s dx=\bm{J}^s(x_1)
\end{equation}

\begin{figure}
        \includegraphics[width=16cm]{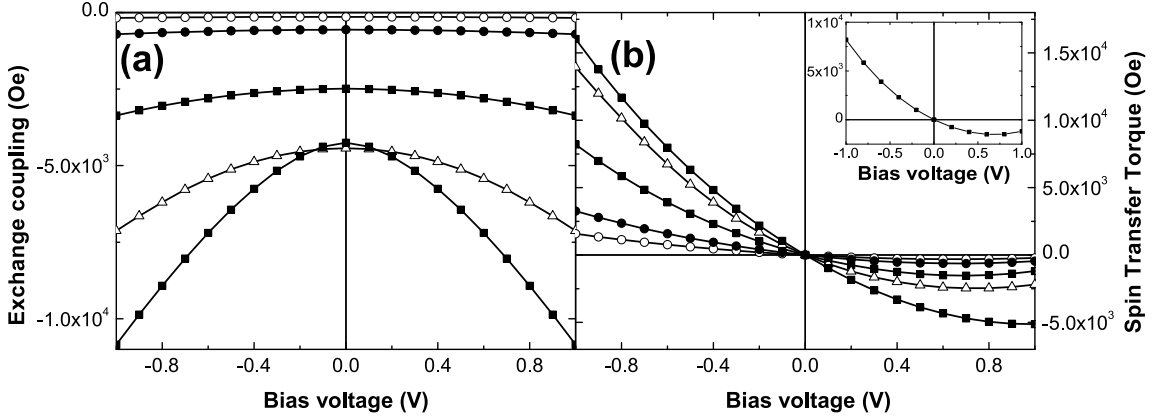}
\caption{Bias dependence of out-of-plane (a) and in-plane (b)
torques for different values of {\it s-d} coupling: $J_{sd}=0.38$
eV (open circles), $J_{sd}=0.76$ eV (filled circles),
$J_{sd}=1.62$ eV (open squares), $J_{sd}=2.29$ eV (open
triangles), $J_{sd}=2.97$ eV (filled squares). Top inset: Bias
dependence of STT for $J_{sd}=1.62$ eV; the solid line was
calculated following the usual way and the symbols were calculated
using Eq. \ref{eq:js2}.} \label{fig:STT_V_Jsd}\end{figure}

Fig. \ref{fig:STT_V_Jsd} displays the total out-of-plane (a) and
in-plane (b) torques as a function of the applied bias voltage,
for different values of the {\it s-d} exchange parameter $J_{sd}$.
Consistently with Theodonis et al. \cite{theo}, the out-of-plane
torque is quadratic whereas the in-plane torque is a combination
between linear and quadratic bias dependence.\par

Finally, note that a change of sign of spin transfer torque at high
positive bias voltage is expected\cite{theo}. The in-plane torque
change of sign should be observed in MTJ with low enough barrier
height and high breakdown voltage (MgO seems a good candidate).
Nevertheless, more technological development are needed to fabricate
such junctions.\par

\begin{figure}
        \includegraphics[width=16cm]{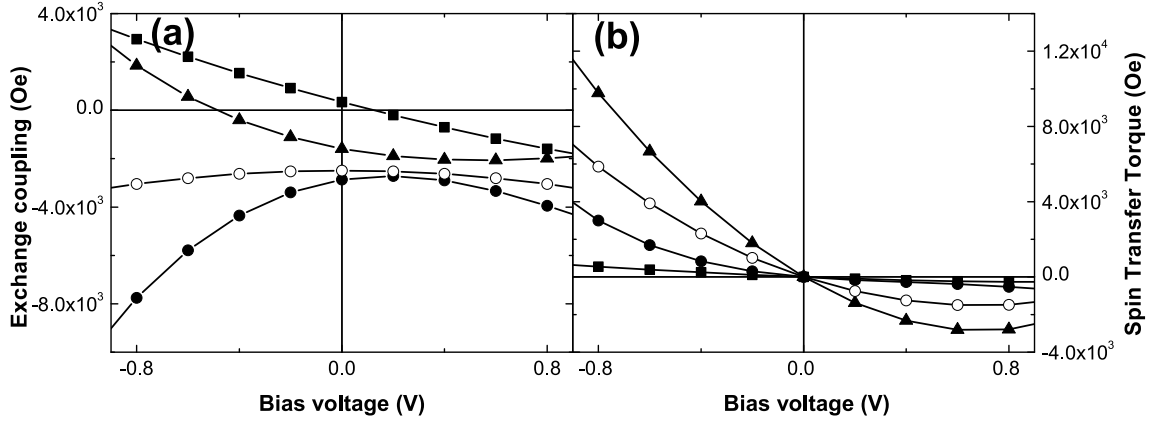}
\caption{Bias dependence of out-of-plane (a) and in-plane (b)
torques for $J_{sd}=1.62$ eV and different values integration
depth: $t=0$ \AA (open squares), $t=4$ \AA (filled triangles),
$t=10$ \AA (filled circles), $t=\infty$ \AA (open
circles).}\label{fig:fig11}
\end{figure}

However, Eq. \ref{eq:js} assumes that all the transverse spin
density is relaxed within the free layer. In other words, the
initially misaligned incident electron spin eventually aligns on
the local magnetization within the free layer. This assumption
seems to be valid, regarding the previous discussions.
Nevertheless, considering weak spin-diffusion processes as well as
non-half metallic junctions (i.e. not like Fe/MgO/Fe), one may
assume that the electron spin is not fully aligned on the local
magnetization when leaving the free layer. This assumption may be
valid in magnetic semiconductor-based tunnel junctions, where the
spin-diffusion length is very large\cite{asga}. Fig.
\ref{fig:fig11} displays the bias dependence of out-of-plane and
in-plane torques for different integration depths $t$ (namely,
different layer thicknesses):

\begin{equation}\label{eq:js23}
    \overrightarrow{T}_{partial}=\int_{x_1}^{x_1-t}-\nabla \bm{J}^s dx=\bm{J}^s(x_1)-\bm{J}^s(x_1-t)
\end{equation}
The bias dependence can change drastically and the out-of-plane
torque can even change its sign (note that the in-plane torque keeps
its general shape). These dependencies are strongly affected by the
tunnel barrier characteristics and one should be careful in the
analysis of bias dependence.\par

\subsubsection{Circuit theory}
\begin{figure}[ht]
\centering
    \includegraphics[width=10cm]{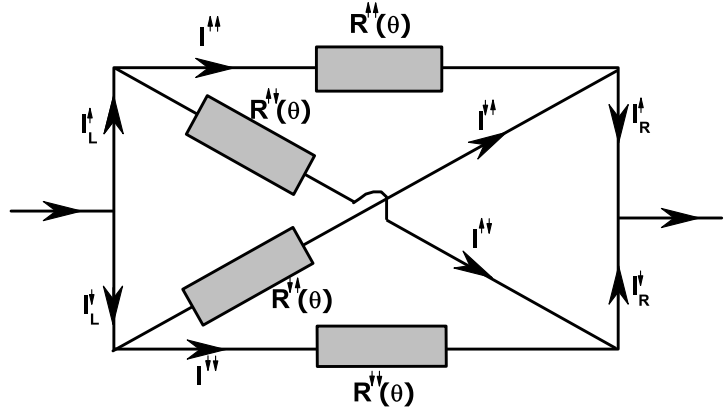}
    \caption{Schematics of the circuit model proposed by Slonczewski \cite{slonc05,slonc02}.}\label{fig:circuit}
\end{figure}

Slonczewski \cite{slonc05} proposed a circuit model to describe
magnetic tunnel junctions in the general case, without restriction
of the band structure of the electrodes and of the barrier. Fig.
\ref{fig:circuit} shows the schematics of this model. Theodonis et
al. \cite{theo} have demonstrated that this model reproduces well
the bias dependence of the in-plane torque. If one considers the
two pure spin states in the quantification axis of the left
electrode $|\uparrow>_L$ and $|\downarrow>_L$, they can be
decomposed on the eigenstates of the right electrode in the
following manner:
\begin{eqnarray}
|\uparrow>_L=\cos\frac{\theta}{2}|\uparrow>_R+\sin\frac{\theta}{2}|\downarrow>_R\\
|\downarrow>_L=-\sin\frac{\theta}{2}|\uparrow>_R+\cos\frac{\theta}{2}|\downarrow>_R
\end{eqnarray}
where $\theta$ is the angle between the magnetizations of the
electrodes. Then, the probability for an electron spin $\sigma$ in
the left electrode to be observed in a spin projection $\sigma'$
in the right electrode is
$P_{\sigma\sigma'}=|<\sigma|\sigma'>|^2$. The associated
resistances indicated on Fig. \ref{fig:circuit} are inversely
proportional to this probability, thus leading to:
\begin{eqnarray}
R^{\sigma\sigma}(\theta)=R^{\sigma}(0)\cos^{-2}\frac{\theta}{2}\\
R^{\sigma\bar{\sigma}}(\theta)=R^{\sigma}(\pi)\sin^{-2}\frac{\theta}{2}
\end{eqnarray}
Using the expression of in-plane spin transfer torque derived by
Slonczewski \cite{slonc05}:
\begin{equation}
a_j=\hbar(J_{L}^\uparrow-J_{L}^\downarrow+(J_{R}^\downarrow-J_{R}^\uparrow)\cos\theta)/2e\sin\theta
\end{equation}
where $J_{L(R)}\sigma$ is the current density of the spin projection $\sigma$ in L(R) electrode, we then find:
\begin{equation}\label{eq:js2}
a_j=\frac{J^s_{AP}-J^s_{P}}{2}
\end{equation}
where $J^s_{AP(P)}$ are the interfacial spin current densities
when the magnetizations are in antiparallel (parallel)
configuration. Theodonis et al. \cite{theo} claimed that this
relation is independent of the electronic structure or of the
adopted description (free electron, tight-binding...). As a matter
of fact, the insert of Fig. \ref{fig:STT_V_Jsd} shows the STT
calculated using Eq. \ref{eq:js} (solid line) and using Eq.
\ref{eq:js2} (symbols), which are in very good agreement. From
Brinkman's model \cite{brink}, the authors demonstrated that the
component $T_{||}$ is the superposition of a linear contribution
$J^s_{P}$ and a quadratic contribution $J^s_{AP}$ as a function of
the bias voltage.\par As a matter of fact, Brinkman et al.
\cite{brink} have showed, from a free electron model, that the
current density flowing across a non magnetic tunnel junction
whose barrier is asymmetric and submitted to a bias $V$ may be
described by:
\begin{eqnarray}
&&J(V)=f_1(\bar{\Phi})V-f_2(\bar{\Phi})\Delta\Phi V^2+O(V^3)\label{e:br}\\
&&\bar{\Phi}=(\Phi_l+\Phi_r)/2\\
&&\Delta\Phi=\Phi_l-\Phi_r
\end{eqnarray}
where $\Phi_l$ and $\Phi_r$ are the barrier height at the left and
right interfaces, measured from the bottom of the conduction band.
$f_1$ and $f_2$ are determined in Ref. \cite{brink}. In the case
of a magnetic tunnel junction, Eq. \ref{e:br} apply to each spin
projection. When the magnetizations are parallel, the MTJ behaves
like a symmetric tunnel junction for each spin projection and
$\bar{\Phi}^\uparrow \neq \bar{\Phi}^\downarrow$,
$\Delta\Phi^\uparrow=\Delta\Phi^\downarrow=0$. On the contrary, if
the electrode magnetizations are antiparallel, the MTJ behaves
like a asymmetric tunnel junction for each spin projection and
$\bar{\Phi}^\uparrow=\bar{\Phi}^\downarrow$,
$\Delta\Phi^\uparrow=-\Delta\Phi^\downarrow$. The spin density is
then:
\begin{eqnarray}
&&J^s_{P}=(f_1(\bar{\Phi}^\uparrow)-f_1(\bar{\Phi}^\downarrow))V+O(V^3)\\
&&J^s_{AP}=-2f_2(\bar{\Phi})V^2+O(V^3)
\end{eqnarray}
By this way, Theodonis et al. \cite{theo} demonstrated that the
general form of the Slonczewski term is $a_j=a_1V+a_2V^2+O(V^3)$.
The balance between the two bias dependencies, quadratic and
linear, may be modified by varying $J_{sd}$.\par

Note that the circuit model cannot describe the second component
$b_j$ of the spin transfer, since it makes two restrictive
assumptions: i) during the transport, the electron spin remains in
the magnetization plane ($\eta=0$ - see Fig.
\ref{fig:anglespin_Jsd}) and ii) the spin current is completely
absorbed at the interface (no precession is taken into account,
since the electron spin is instantaneously reoriented along the
local magnetization). These two hypothesis ignore the effects
which give rise to the out-of-plane torque\cite{slonc05}.\par

\subsubsection{Asymmetric junction}

Wilczynski et al. \cite{barnasstt} recently showed that the bias
dependence of the torque is strongly affected by the symmetry of
the junction. Considering two different ferromagnetic electrodes
(different thickness or different {\it s-d} exchange coupling),
the authors show that the bias dependence may be very different
from the usual parabolic and second order bias dependence depicted
in Fig. \ref{fig:STT_V_Jsd}.\par Slonczewski et al. \cite{slonc07}
recently proposed a study of the influence of elastic and
inelastic tunnelling in the spin transfer torque characteristics.
This discussion is restricted to the in-plane torque and the
out-of-plane component is predicted to be in the second order of
bias voltage.

\subsubsection{Role of magnons emissions}

Magnons emission are also expected to play an important role in
spin-dependent tunnelling transport. As a matter of fact, Zhang et
al. \cite{zhang97} proposed that impinging electrons with energy
higher than the Fermi level can emit spin waves by flipping their
spin near the MTJ interface, leading to TMR drop as a function of
the applied bias voltage. Levy and Fert\cite{swmtj} recently
suggested that the partial depolarization of spin-current by
spin-waves emission may give rise to a torque on the local
magnetization, and consequently significantly contribute to spin
transfer torque. We give here a summary of the picture proposed in
Ref. \cite{swmtj}.\par

The authors considered a system similar to Slonczewski's
\cite{slonc89} where the barrier is rectangular and submitted to
low bias voltage. In this case, we saw that only in-plane spin
transfer torque appears (see Eq. \ref{e:slonc}). The authors
showed that in the case of spin-waves emission, the in-plane
torque possesses four sources:
\begin{equation}
\bm{T}_{||}=(T^{elas}+T^{int}+T^{bulk trans}+T^{bulk
long})\textbf{M}\times(\textbf{M}\times\bm{P})
\end{equation}
where the four terms stand for the elastic torque (usual in-plane
torque), the emission of interfacial magnons and the emission of
bulk magnons acting on the transversal and longitudinal component of
the local magnetization.\par\medskip

\paragraph{Interfacial magnons}
Magnons in general can only be excited by electrons whose energy is
higher than the Fermi level and, their energy is $\hbar
\omega^{l(r)}_q<eV$. This leads to the formulation of the torque due
to interfacial magnons excitations, exerted on the $left$ layer:
\begin{eqnarray}
T^{int}_l\propto|t^i|^2\sin\theta
V^2\{\alpha_rN^i_lP_r+N^i_r(P_l\cos\theta+F(\theta))\}\nonumber
\end{eqnarray}
where $N^i_{l(r)}$ are the numbers of spins per unit area at the
interface (in the left and right electrodes, respectively),
$P_{l(r)}$ are the interfacial spin polarizations, $\alpha_{l(r)}$
are coefficients which include material parameters and $F(\theta)$
is a function of $\theta$ that we do not define here (see Ref.
\cite{swmtj}). This form is complex and shows quadratic dependence
as a function of the bias voltage. Furthermore, the authors found
that the torques induced by interfacial magnon emission, applied
to left and right electrode, are in opposite direction (favors
parallel alignment of the left magnetization and antiparallel
alignment of the right magnetization).
\begin{equation}
T^{int}_r=-T^{int}_l(l\longleftrightarrow r)
\end{equation}
To understand this effect, Levy and Fert\cite{swmtj} give the
following argument. The elastic spin current polarization arises
from the weighted contribution of both left and right magnetic
electrodes.\par

For the electrode at the higher electrochemical potential, left
electrode here, the authors found that the magnon emitted in this
electrode causes the polarization to shift toward the polarization
of the right electrode, which effectively is in the same direction
than elastic torque.\par

However, for the electrode with the lower electrochemical potential,
right electrode here, this reorientation of the polarization reduces
the effect of the elastic term, creating an additive torque in the
opposite direction.\par\medskip

\paragraph{Bulk magnons}

Considering the electrons which kept their spin close to the
interface, one has to distinguish between two behaviors. Some of
these electrons are scattered with spin-flip in the bulk magnetic
lead whereas others are scattered without spin-flip. The spin-flip
scattered electrons contribute to a {\it transverse} component of
the spin current. This leads to the torques due to bulk magnon
emission, exerted on the left and right electrodes:
\begin{eqnarray}
&&T^{bulk\;trans}_l\propto V^{3/2}|t^b|^2\sin\theta N^b_r\left[P_l\cos\theta+F'(\theta)\right]\\
&&T^{bulk\;trans}_r\propto V^{3/2}|t^b_m|^2\sin\theta N^b_r
\end{eqnarray}
where $N^b_{l(r)}$ are the numbers of spins per unit volume. The
electrons scattered without spin flip also contributes to the
torque, by affecting the {\it longitudinal} component of the spin
current. When incoming in the right electrode, they do not
contribute to the torque on this electrode, but this reduction of
the longitudinal part of the spin current contributes to a torque on
the left magnetic lead.
\begin{eqnarray}
&&T^{bulk\;long}_l\propto V^{3/2}|t^b_m|^2\sin\theta\cos\theta N^b_r\\
&&T^{bulk\;long}_r=0
\end{eqnarray}
This study suggests that the torque due to magnon emission by hot
electrons arises from 4 different mechanisms, and has a
self-consistent form. The authors used this theory to explain the
data gathered by Fuchs et al. \cite{fuchs0} (see section
\ref{s:sttmr}). We stress out that this model is restricted to low
bias voltage and the authors point out that other factors may
influence spin torque properties such as the energy dependence of
the interfacial density of states, which was considered in
Theodonis et al. \cite{theo}, Wilczynski et al. \cite{barnasstt}
and Manchon et al. \cite{manchon} theories.

\subsection{Recent experimental investigations}

As discussed in section \ref{s:torques}, a number of experiments
have been carried out in order to determine the characteristics of
sin transfer torques in magnetic tunnel junctions. Early
experimental studies by Fuchs et al. \cite{fuchs06} demonstrated a
linear variation of in-plane torque as a function of the applied
bias voltage. However, no determination of the out-of-plane
component was reported until the publication of very recent
experiments.\par These experiments are of two types. The first
ones use radio-frequency techniques, addressing FMR or magnetic
noise under spin torque, while the second ones use the quasistatic
stability phase diagrams to describe spin torque properties.

\subsubsection{Radio-frequency signature of spin torque}

The spin-diode effect studied by Tulapurkar et al. \cite{tula} was
firstly explained using a linear bias dependence for the two terms
of spin torque, $a_j$ and $b_j$, consistently with the first study
of Petit et al. \cite{seb} concerning the influence of spin torque
in thermally activated FMR excitations. Although this
interpretation has now been questioned by recent experiments,
these studied demonstrated the necessity to take into account an
out-of-plane component of the torque in order to interpret the
experimental results.\par

The very recent studies of Sankey et al. \cite{sankeynat} and
Kubota et al. \cite{kubota} constitute a breakthrough in the
experimental determination of spin torque since the authors were
able to reconstruct the bias dependence of both torque components
by fitting the experimental results (note that Sankey et al.
\cite{sankeynat} give the "torkance"\cite{slonc07} bias
dependence).

\begin{figure}[ht]
    \centering
        \includegraphics[width=8cm]{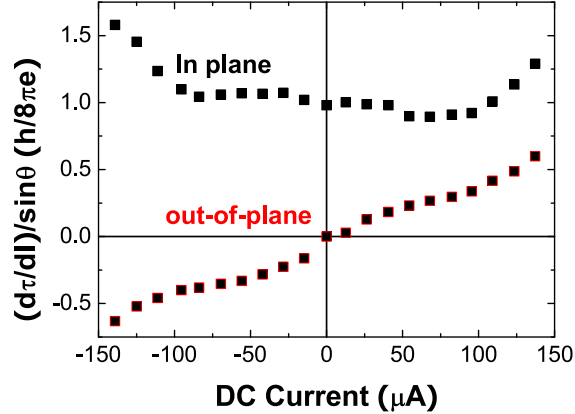}
\caption{Bias dependence of torkance for the in-plane and
out-of-plane torques. From Ref. \cite{sankeynat}.}
    \label{fig:sankey}
\end{figure}

Both studies prove a quadratic bias dependence of the $b_j$ term
as well as a second order polynomial dependence of $a_j$ (see Fig.
\ref{fig:sankey}), confirming the recent theories on spin torque
in MTJ\cite{theo,manchon,slonc07}. Furthermore, both torques are
found to be of the same order of magnitude.\par
The determination of the bias dependence of the out-of-plane
component is very tricky since this torque only induces a small
shift in the resonance peaks of the measured signals. Furthermore,
the treatment of temperature issues (temperature dependence of the
signal, thermal activation, Joule effects, Peltier effects and
"thermal spin transfer torque"\cite{hatami}) as well as de-embedding
procedure must be properly undertaken.

\subsubsection{Thermally activated phase diagrams}

Very recent experiments, not yet published, have proposed to study
the thermally activated phase diagrams of magnetic tunnel junctions
in order to describe the spin transfer torque bias dependence. Such
phase diagram shows the stable magnetic state of the free layer of a
spin-valve device, as a function of both the applied field and the
injected current.\par
 A first experiment was performed by Li et
al. \cite{prlli} in order to get the bias dependence of torques
from the bias dependence of the critical switching fields of the
free layer of a MgO-based MTJ. The authors used short bias voltage
pulses to increase the maximum bias voltage above the quasistatic
breakdown voltage without damaging the junction. They succeeded in
describing the in-plane and out-of-plane torques, claiming a
linear bias dependence for the first and a mostly quadratic
dependence for the second one. However, contrary to previous
results, the authors give a bias dependence of the form
$b_j\propto VJ$, where $J$ is the current density flowing through
the junction.\par

\begin{figure}
    \centering
        \includegraphics[width=9cm]{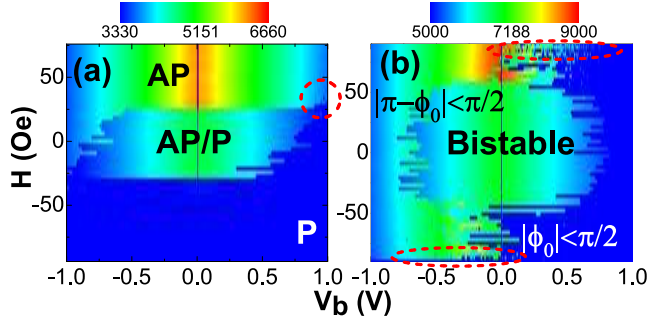}
\caption{Static phase diagrams of magnetic tunnel junction with
longitudinal (a) [Sample A] and transverse applied field (b)
[Sample B]. The red circles show the magnetic excitation regions.
The color code refers to the resistance of the
stack}\label{fig:fig1}
\end{figure}

Manchon et al. \cite{manchonkj} used a slightly different
technique, without short pulses and succeeded to draw a complete
phase diagram in two different magnetic configurations: (a) when
the external field is applied along the easy axis of the free
layer and (b) when the external field is applied along the hard
axis of the free layer. These diagrams are given in Fig.
\ref{fig:fig1}, for two different samples (A and B).\par
\begin{figure}
    \centering
        \includegraphics[width=9cm]{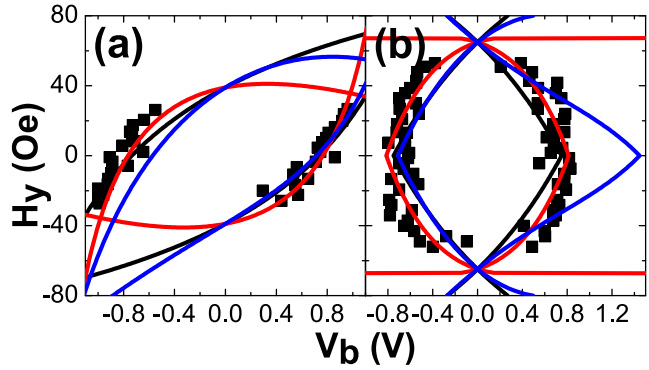}
\caption{Analytical fits of the critical lines (symbol) of
longitudinal (a) and transverse (b) static phase diagrams, with
$b_j=0$ (black), $b_j=b_2V^2$ (blue) and $b_j=b_2V$
(red).}\label{fig:fig2bis}
\end{figure}

Assuming, in a first approximation, that the in-plane torque is
linear as a function of bias voltage, several fits of the
thermally activated phase diagrams were performed, using the
theory of thermal activation developed by Koch et al.
\cite{koch,Sun,Li}. Fig. \ref{fig:fig2bis} shows the three fits
the authors obtained, assuming $b_j=0$ (black), $b_j=b_2V^2$
(blue) and $b_j=b_2V$ (red), where $b_2$ is a fitting
parameter.\par
Assuming a quadratic bias dependence of the out-of-plane torque
term introduces an significant asymmetry in both longitudinal and
transverse stability diagrams that is not observed experimentally.
Furthermore, although no significant difference appears in the
transverse stability diagram when assuming $b_j$ = 0 or $b_j= b2V$
(black and red curves in Fig. \ref{fig:fig2bis}(b)), the best fit
of the longitudinal diagram is clearly obtained when  $b_j$ is
linear. This indicates that in our samples,  $b_j$ should be an
odd function of the applied bias V , contrary to Sankey et al.
\cite{sankeynat} and Kubota et al. \cite{kubota}.\par

This linear bias dependence is in contradiction with the recent
published theories\cite{theo,manchon} predicting a quadratic bias
dependence of the out-of-plane torque. These theories assume
amorphous tunnel barrier, low bias voltage, semi-infinite free layer
thickness and zero temperature whereas we performed our measurements
on MTJs comprising crystalline MgO barrier at room temperature.
Consequently, the differences between our experiments and these
theories may be ascribed to the crystalline nature of the MgO
barrier as well as other contributions such as spin-waves emissions
that have not been considered in the calculations despite their
strong influence on the spin torque bias dependence\cite{swmtj}.\par

The difference with the recent RF
measurements\cite{sankeynat,kubota} are more difficult to interpret.
It may be attributed to the interplay between thermal effects and
current-induced magnetization dynamics. Note that the results
obtained by RF measurements strongly depend on the samples
quality\cite{suzuki} and may present a linear $b_j$ term.\par

These experiments are of great interest because of its relative
simplicity. However, further experimental improvements are needed in
order to increase the reproducibility and accuracy of the
measurements and be able to measure both longitudinal and transverse
phase diagram on the same sample without breakdown.

\section{From weak ferromagnetic to half-metallic tunnel junctions\label{s:wfhm}}

To conclude this chapter, we studied the dependence of the
in-plane and out-of-plane torque as a function {\it s-d} exchange
coupling $J_{sd}$, and in particular, the crossover between
ferromagnetic and half-metallic tunnel junctions. As a matter of
fact, as previously stated, Heiliger et al. \cite{heiliger}
suggested that a crystalline MgO-based tunnel junction may be
approximated by a half-metallic tunnel junction, when considering
the dominant contribution of $\Delta_1$ symmetry.\par

\begin{figure}
    \centering
        \includegraphics{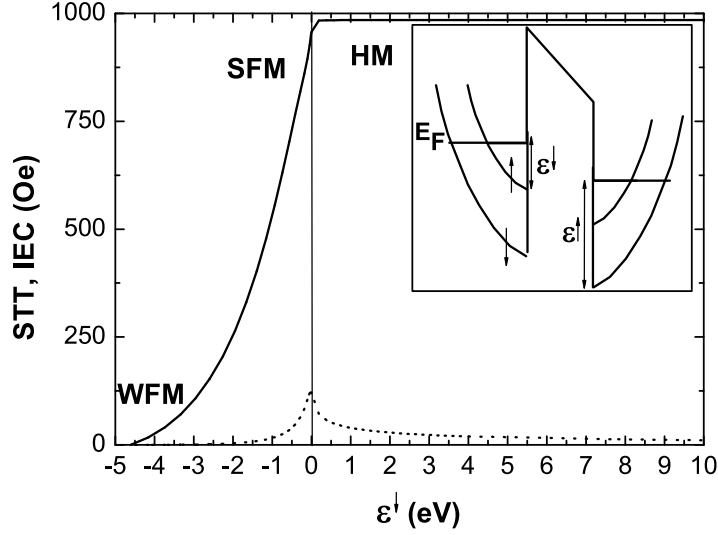}
\caption{In-plane (solid line) and out-of-plane (dotted line)
torques as a function of {\it s-d} exchange coupling. The vertical
line shows the limit between ferromagnetic (weak ferromagnetic
-WFM- and strong ferromagnetic -SFM-) regime and half-metallic
regime.} \label{fig:fig12}
\end{figure}

The Fermi energy is kept constant, whereas the energy of the
bottom of the minority electrons conduction band
$\epsilon^{\downarrow}$ is modified, as indicated in Fig.
\ref{fig:fig12}. This energy is defined from the Fermi energy as:
\begin{equation}
    \epsilon^{\downarrow}=E_F-E_c^{\downarrow}=-\frac{\hbar^2k_F^{\downarrow2}}{2m}
\end{equation}
where $E_c^{\downarrow}$ is the absolute energy of the bottom of
the conduction band. When $\epsilon^{\downarrow}$ is close to
$\epsilon^{\uparrow}$, $k_F^{\uparrow}\approx k_F^{\downarrow}$,
the metallic electrodes loose their ferromagnetic nature. For
$\epsilon^{\downarrow}\approx 0$, the Fermi wavevector for
minority electrons becomes smaller and the current polarization is
strongly enhanced. In this case, we expect an important spin
transfer torque. When $\epsilon^{\downarrow}>0$,
$k_F^{\downarrow}$ becomes imaginary and the electrodes behave
like a tunnel barrier for minority spins. Increasing
$\epsilon^{\downarrow}$ increases the evanescent decay of minority
wave functions in the electrodes. Then, the product
$<\Psi^{*\uparrow}\Psi^{\downarrow}>$ still exists so that spin
torque is non zero and decrease exponentially from the
interface.\par

Fig. \ref{fig:fig12} shows the amplitude of in-plane and
out-of-plane torques in the three different regimes: weak
ferromagnetic electrodes (WFM), strong ferromagnetic electrodes
(SFM) and half-metallic electrodes (HM). As expected, in
ferromagnetic regime, in-plane and out-of-plane torques increase
until $\epsilon^{\downarrow}=0$ (vertical line). When
$\epsilon^{\downarrow}$ becomes positive, the bottom of the
conduction band of minority electrons lies above the Fermi level:
no minority electrons can propagate because only evanescent states
exist near the interfaces for this spin projection. However,
in-plane and out-of-plane torques do not vanish but reach a
plateau which slowly decreases to zero when increasing $J_{sd}$
(not shown).\par
\begin{figure}
    \centering
        \includegraphics{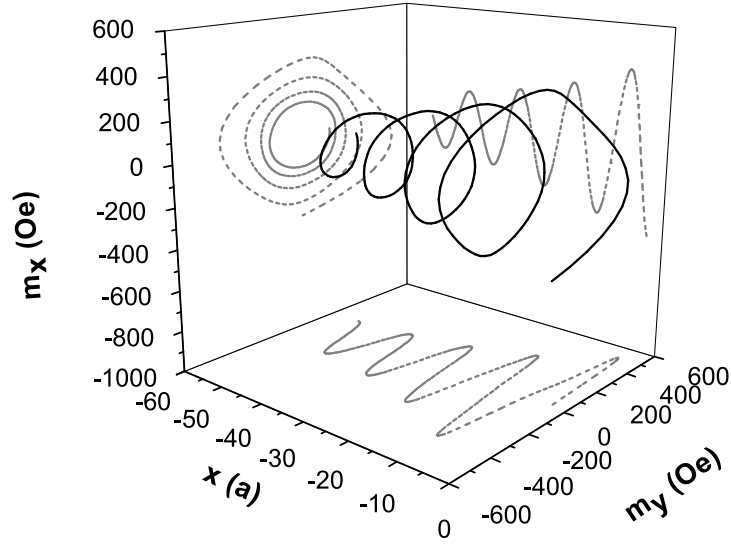}
\caption{Transverse spin density (black line) as a function of the
penetration depth from the barrier within the left ferromagnetic
electrode in a usual ferromagnetic regime. We set
$\epsilon^\downarrow=-1.37$ eV and $V_b=0.1$ V.}\label{fig:fig13}
\end{figure}

\begin{figure}
    \centering
        \includegraphics{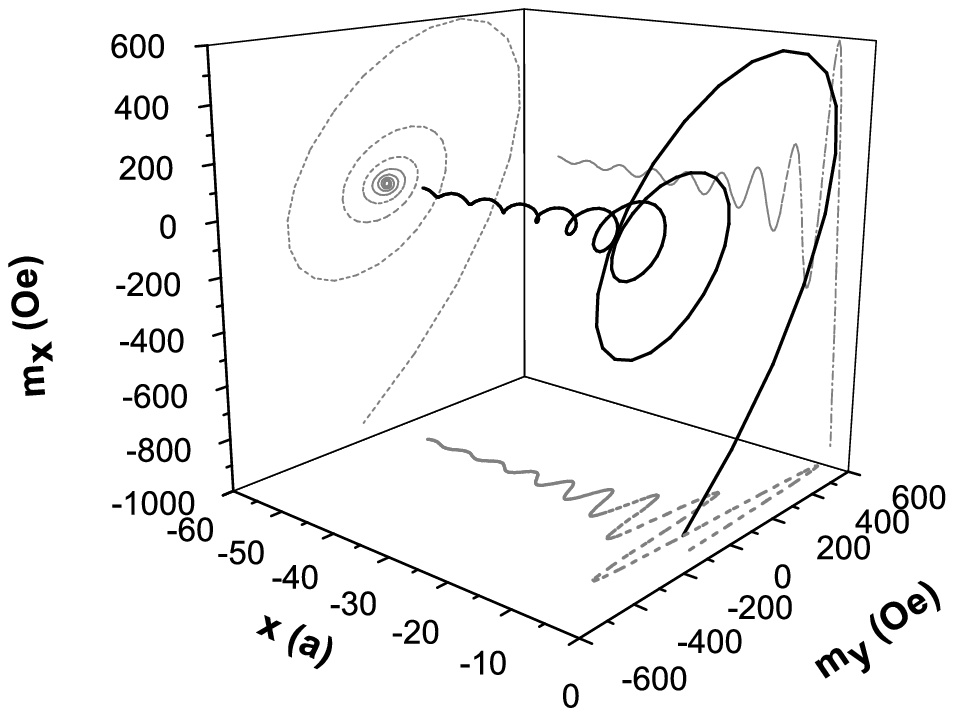}
\caption{Transverse spin density (black line) as a function of the
penetration depth from the barrier within the left ferromagnetic
electrode in a strong ferromagnetic regime. We set
$\epsilon^\downarrow=-0.38$ eV and $V_b=0.1$ V.}\label{fig:fig14}
\end{figure}

To understand this behavior, we calculated the spatial dependence
of the transverse spin density in the free layer. Fig.
\ref{fig:fig13} shows the transverse spin density in a usual
ferromagnet, $\epsilon^\downarrow=-1.37$ eV (which corresponds to
$J_{sd}=1.62$ eV), as a function of the distance from the
interface with the barrier in the left electrode. The oscillation
possesses the same characteristics than discussed above and we
observe that the transverse spin density is damped far from the
interface. When decreasing $\epsilon^\downarrow$, the interfacial
spin density increases, due to strong spin filtering at the
interface (strong spin-dependent selection), as shown on Fig.
\ref{fig:fig14}.\par
\begin{figure}
    \centering
        \includegraphics{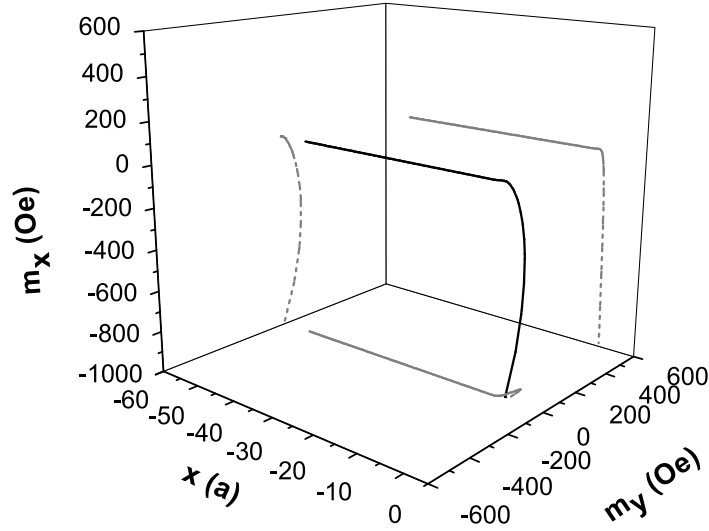}
\caption{Transverse spin density (black line) as a function of the
penetration depth from the barrier within the left ferromagnetic
electrode in half-metallic regime. We set $\epsilon^\downarrow=19$
eV and $V_b=0.1$ V.}\label{fig:fig15}
\end{figure}

But when $\epsilon^\downarrow$ changes sign, only majority electrons
can propagate and the transverse spin density becomes:
\begin{eqnarray}
&& m_x^\uparrow=16q_1q_2\sin\theta\ \Re\{(k_3-k_4)\left(\frac{e^{-i(k_1+k_2)(x-x_1)}-r_1^{*\uparrow}e^{i(k_1-k_2)(x-x_1)}}{den}\right)\}\\
&& m_y^\uparrow=-16q_1q_2\sin\theta\ \Im\{(k_3-k_4)\left(\frac{e^{-i(k_1+k_2)(x-x_1)}-r_1^{*\uparrow}e^{i(k_1-k_2)(x-x_1)}}{den}\right)\}
\end{eqnarray}
where $q_{1,2}$ are the barrier wave vectors at the left and right
interface respectively, $k_{1,2}(k_{3,4})$ are the electron wave
vectors in the left (right) electrode for majority and minority
spins, respectively, and $den$ is a coefficient which depends on the
junction parameters. Considering Fermi electrons at perpendicular
incidence, very small bias voltage ($eV\approx0$) and imaginary
minority electron spin wave vector, $k_{2(4)}=ik$, we obtain
straightforwardly:
\begin{eqnarray}
&& m_x^\uparrow=16q_1q_2 e^{k(x-x_1)}\sin\theta\ \Re\{(k_3-ik)\left(\frac{e^{-ik_1(x-x_1)}-r_1^{*\uparrow}e^{ik_1(x-x_1)}}{den}\right)\}\\
&& m_y^\uparrow=-16q_1q_2e^{k(x-x_1)}\sin\theta\ \Im\{(k_3-ik)\left(\frac{e^{-ik_1(x-x_1)}-r_1^{*\uparrow}e^{ik_1(x-x_1)}}{den}\right)\}
\end{eqnarray}
The transverse spin density is a product between oscillating
function of $k_1$ and exponentially decaying function of $k$. Fig.
\ref{fig:fig15} shows the spatial evolution of the transverse spin
density in the case of a half-metallic tunnel junction. All the
oscillations are damped very quickly so that the only important
contribution to torque comes from the interface. Contrary to usual
MTJ (where both bulk averaging due to spatial interferences and
interfacial spin reorientation contribute to spin torque), in a
strong half-metallic tunnel junction all the torque comes from
spin reorientation due to spin-dependent reflection. In this last
case, the contribution of the spatial averaging between all
impinging electrons ($\kappa$-summation) is reduced compared to
interfacial spin transfer.\par
The interesting point is that half-metallic tunnel junctions may
reproduce the general properties of MgO-based tunnel junctions.
Most of the previous characteristics discussed earlier (quantum
description as well as observable characteristics) are then valid
in this type of junctions. This explains why simple single band
per spin models, like the one proposed by Theodonis et al.
\cite{theo} for simple cubic crystal structure, or Manchon et al.
\cite{manchon}, assuming amorphous tunnel barrier, applies to
experimental results obtained in crystalline MgO-based MTJ. Note
however that this agreement holds for thick enough MgO barriers
and that the quality of the tunnel junction should strongly affect
the half-metallic characteristic. Other symmetry channels may then
contribute to the transport, like resonant interfacial states for
example \cite{tsymbal,tiusan07}.\par

Kubota et al. \cite{kub2006} recently studied the dependence of
the critical switching current density on the thickness of the
free layer in a MgO-based MTJ. The authors found that the critical
current density was roughly proportional to the free layer
thickness. This indicates that the transverse spin current is
completely absorbed within the free layer, and that consequently
the spin transfer torque seems to take place close to the
interface between the insulator and the ferromagnetic electrode,
consistently with the above discussion.

\section{Conclusion}

As stated in the introduction, since its first prediction
\cite{slonc89} and observation \cite{huai,fuchs}, spin transfer
torque in tunnel junctions was expected to present strong
differences compared to spin torques in metallic spin valves. The
single-electron nature of the tunnelling transport, the specific
spin-selection induced by the tunnel barrier, as well as the non
linearity of the tunnelling process itself were expected to
strongly affect the observable properties of spin transfer
torque.\par

The smaller role of spin accumulation is also of great importance
since the angular dependence of spin torque coefficient $a_j$ and
$b_j$ are unusually small in MTJs. Another characteristic is the
significant amplitude of the out-of-plane component of spin
transfer torque, arising from the spin-selection occurring at the
tunnel barrier.\par

Most interesting, recent experiments based on RF techniques or
(quasi-)static measurements have revealed significant non
linearities in the spin torque bias dependence, due to the
non-linearity of the tunnelling transport. The most striking element
is that these experiments seem to agree with tight-biding or
free-electron models, i.e. models making very simplistic and
restrictive assumptions on the energy dependence of the interfacial
densities of states and on the barrier shape. Although it has been
widely shown that MgO-based tunnel junctions possess a complex
electronic band structure, these experiments are conveniently
reproduced by parabolic or bell-like band structure. This surprising
simplicity may be attributed, as proposed in section \ref{s:wfhm},
by the dominant contribution of $\Delta_1$ symmetry electrons, at
low bias and not-too-thin barrier width.\par

However, more accuracy is needed both in the theories and
experiments in order to better describe these specificities.
Junctions asymmetries, inelastic scattering or impurities have been
shown to deeply modify the spin torque properties in MTJs.
Hot-electrons spin-waves emission is also known to be of great
importance in MTJs, leading to the so-called "zero-bias anomaly".
This emission is also expected to significantly affect the bias
dependence of spin transfer torque.\par

We stressed out the simplicity of the models that have been proposed
up to now to describe spin torques in MTJs. Realistic band structure
calculations should enrich our knowledge of spin torque origins,
especially by modifying the spin-filtering mechanism and the
interference process between the majority and minority electrons.
The ballistic assumption, namely neglecting all spin-flip
scattering, limits the investigation to academic systems. Taking
spin-orbit coupling into account would be of great interest to
quantitatively simulate real magnetic devices.\par

Finally, nothing have been said in this chapter about the
time-domain investigations of magnetization dynamics in MTJs.
Preliminary experimental studies were carried out by Devolder et
al. \cite{devolder} that show interesting magnetic behaviors not
observed in metallic spin-valves until now.\par

As we tried to show in this chapter, although quite incomplete,
the recent research on spin transfer in MTJs has already revealed
rich and exciting issues that only wait for further theoretical
and experimental efforts.\clearpage

\begin{acknowledgments}
The authors acknowledge fruitful discussions with Prof.
Zhang, Prof. Kubota, T. Devolder, Prof. Miltat, M.D. Stiles. We
gratefully acknowledge the support of J.-E. Lee and S.-C. Oh from
Samsung Electronics Co. for providing the MgO-based samples. We
also thank Prof. D.C. Ralph, G. Fuchs and S. Petit for their
agreement in reproducing apart of their exciting work.\par

Financial support by the European Commission within the EU-RTNs
SPINSWITCH (MRTN-CT-2006-035327) is gratefully acknowledged. The
views expressed are solely those of the authors, and the other
Contractors and/or the European Community cannot be held liable for
any use that may be made of the information contained herein.\par

This work was partially supported by the Russian fund for basic
research. \end{acknowledgments}

\end{document}